\documentclass[review,onefignum,onetabnum]{siamart190516}
\usepackage{soul} 

\usepackage{xcolor,cite}

\usepackage{amssymb}
\usepackage[many]{tcolorbox}
\newtcolorbox{boxE}{
    sharpish corners, 
    boxrule = 0pt,
    fuzzy shadow = {0pt}{-2pt}{-0.5pt}{0.5pt}{black!75} 
}

\newcommand{\ep}{\epsilon}
\newcommand{\mc}{\mathcal}

\usepackage{lipsum}
\usepackage{amsfonts}
\usepackage{graphicx}
\usepackage{epstopdf}
\usepackage{algorithmic}
\ifpdf
  \DeclareGraphicsExtensions{.eps,.pdf,.png,.jpg}
\else
  \DeclareGraphicsExtensions{.eps}
\fi

\newsiamremark{remark}{Remark}
\newsiamremark{hypothesis}{Hypothesis}
\crefname{hypothesis}{Hypothesis}{Hypotheses}
\newsiamthm{claim}{Claim}

\headers{Short term plasticity alters wandering bump dynamics}{H.L. Cihak and Z.P. Kilpatrick}

\title{Multiscale motion and deformation of bumps in stochastic neural fields with dynamic connectivity
\thanks{Submitted to the editors June 27, 2023\funding{This work was funded by a Collaborative Research in Computational Neuroscience grant, NSF-DMS-2207700, and a BRAIN Initiative grant, NIH-R01EB029847.}}}

\author{Heather L Cihak\thanks{Department of Applied Mathematics, University of Colorado Boulder, Boulder, CO 
  (\email{Heather.Cihak@colorado.edu}, \email{zpkilpat@colorado.edu})}
\and Zachary P Kilpatrick\footnotemark[2]
}

\usepackage{amsopn}

\usepackage{xr}
\makeatletter
\newcommand*{\addFileDependency}[1]{
  \typeout{(#1)}
  \@addtofilelist{#1}
  \IfFileExists{#1}{}{\typeout{No file #1.}}
}
\makeatother

\usepackage[margin=0.5in]{geometry}

\begin{document}
\nolinenumbers

\maketitle
\begin{abstract}
 The distinct timescales of synaptic plasticity and neural activity dynamics play an important role in the brain's learning and memory systems. Activity-dependent plasticity reshapes neural circuit architecture, determining spontaneous and stimulus-encoding spatiotemporal patterns of neural activity. 
 Neural activity bumps maintain short term memories of continuous parameter values, emerging in spatially-organized models with short term excitation and long-range inhibition. Previously, we demonstrated nonlinear Langevin equations derived using an interface method accurately describe the dynamics of bumps in continuum neural fields with separate excitatory/inhibitory populations. Here we extend this analysis to incorporate effects of slow short term plasticity that modifies connectivity described by an integral kernel.
 Linear stability analysis adapted to these piecewise smooth models with Heaviside firing rates further indicate how plasticity shapes bumps' local dynamics. Facilitation (depression), which strengthens (weakens) synaptic connectivity originating from active neurons, tends to increase (decrease) stability of bumps when acting on excitatory synapses. The relationship is inverted when plasticity acts on inhibitory synapses. Multiscale approximations of the stochastic dynamics of bumps perturbed by weak noise reveal the plasticity variables evolve to slowly diffusing and {\em blurred} versions of that arising in the stationary solution. Nonlinear Langevin equations associated with bump positions or interfaces coupled to slowly evolving projections of plasticity variables accurately describe the wandering of bumps underpinned by these smoothed synaptic efficacy profiles.
\end{abstract}

\begin{keywords}
  neural fields, short term plasticity, bump attractor, wave stability, stochastic differential equations, perturbation theory
\end{keywords}

\begin{AMS}
 35Q92, 35B36, 60G07
\end{AMS}


\section{Introduction}
\label{sec:Introduction}
Maintaining information over brief periods of time is an essential component of working memory, a brain function paramount for the completion of daily tasks~\cite{GoldmanRakic1995} and learning. For example, parametric working memory is required to maintain the representation of continuous object features (e.g., position, orientation, color) observed a short time ago or estimated from accumulated evidence~\cite{hulse2020mechanisms,romo1999neuronal,Funahashi89}.  This form of working memory can utilize persistent and spatially localized neural activity, often sustained by a combination of local and feature-specific excitation with long-range or global inhibition~\cite{Constantinidis16,GoldmanRakic1995}, and can also be distributed across multiple brain areas~\cite{Curtis06}. Activity-dependent synaptic plasticity (dynamic connectivity between neural sub-populations) acting on the order of a few seconds has also been proposed to play a role in the accuracy and robustness of maintained memories across time~\cite{mongillo2008synaptic,itskov2011short}, but can also bias estimates across trials~\cite{Kilpatrick18STP_FacSCIREP,barbosa2020interplay}. 

Neural recordings from non-human primates performing visuospatial working memory tasks reveal persistent activity as {\em bumps} when organized according to neurons' feature preference~\cite{Constantinidis16}. In the oculomotor delayed response task, subjects retain a memory of an object's position after it is briefly flashed on a video screen. Cells recorded during this maintenance (delay) period are tuned to object locations so the preference of the most active neurons predicts the response~\cite{Funahashi89,Fuster73}. This activity pattern wanders stochastically during the delay period as a time-dependent degradation of the memory consistent with response errors~\cite{Wimmer14}. Moreover, response biases across trials and resistance to distractors are consistent with physiological processes acting on slow timescales~\cite{lorenc2018flexible}. For instance, short term facilitation, which strengthens synapses originating from active neurons on the order of a few seconds~\cite{tsodyks1998neural}, can attract and stabilize persistent activity to previously stimulated locations~\cite{Seeholzer19,Kilpatrick18STP_FacSCIREP}.

Continuum neural field models support spatiotemporal patterns of neural activity in the dynamics of integrodifferential equations in which the kernel of the integral term describes the polarity and spatial dependence of synaptic interactions~\cite{ermentrout1998neural,bressloff2011spatiotemporal}. Models with weight kernels that are excitatory (E) at short ranges and inhibitory (I) at long ranges produce self-sustaining and marginally stable peaked activity {\em bumps} ~\cite{Amari77,Pinto2001b}. Bumps in both these and other (e.g., spiking) models have been adopted as idealized descriptions of persistent neural activity encoding parametric working memory~\cite{Compte2000}. Their translation symmetry and marginal stability are important linked features which allow for the encoding of a continuum of stimulus feature values while at the same time resulting in fragility to perturbation so that distracting stimuli and dynamic fluctuations degrade memory over time in ways aligned with subject response statistics~\cite{camperi1998model,Wimmer14,bays2015spikes}.

Slow feedback processes have previously been incorporated into neural field models to account for a number of rich spatiotemporal patterns observed in imaging studies~\cite{pinto2001a,kilpatrick2010effects}. Traveling pulses, spiral waves, and self-sustained spatiotemporal oscillations can be accounted for by a combination of synaptic excitation and slow negative feedback attributed to local processes like spike rate adaptation~\cite{huang2004spiral,coombes2003waves,folias2004breathing}. Some studies have carefully identified the impacts of short term plasticity on spatiotemporal dynamics, but typically overlook the intricacies of interactions of distinct and plastic E/I populations~\cite{ermentrout2002regular,york2009recurrent}. In particular, short term facilitation acting on single population models has been shown to stabilize bumps to perturbations~\cite{itskov2011short,Kilpatrick18STP_FacSCIREP}. 
Here we extend our recent analysis of stochastic bump wandering in a neural field model with distinct E and I populations~\cite{cihak2022distinct} to incorporate the effects of activity dependent synaptic plasticity, which operates on a slower timescale.
Short term facilitation (depression) on E synapses tends to attract (repel) bumps, leading to a decrease (increase) in the bump's position variance, while imposing these processes on the I population yields the opposite effect.

Our neural field model is introduced in \cref{sec:TheModels}, incorporating pre-synaptic plasticity mechanisms acting on a slower timescale. Stationary bump solutions and their stability are then analyzed to determine how perturbations distort and shift the bump solution, especially depending on the amplitude and polarity of plasticity~\cref{sec:Deterministic}. Piecewise linear dynamics emerging from the stepped nature of stationary plasticity profiles are treated by partitioning the space of bump perturbations depending on their signs at the edges, extending prior work~\cite{bressloff2011two}. Stochastic dynamics of bumps subject to weak noise are asymptotically approximated by considering how slowly evolving plasticity profiles shift and deform in response to the rapid wandering of E and I bumps (\cref{sec:Stochastic}). The dynamical blurring of plasticity profiles can be approximated via a convolution of the initial profile with a slowly temporally evolving kernel. Our derived nonlinear Langevin equation captures these features in the coupling between neural activity and plasticity variables. Statistics of these reduced-dimension multiscale equations are well-aligned with the results of full simulations across a broad range of parameters, characterizing the short timescale relaxation of the variance to a linear scaling at long times. Further approximations can be made at long timescales, linearizing the Langevin equations to produce a multivariate Ornstein-Uhlenbeck process whose covariance and long term diffusion can be calculated explicitly.


\section{The Model}
\label{sec:TheModels}

\begin{figure}[b!]
  \centering
  \includegraphics[width=0.85\textwidth]{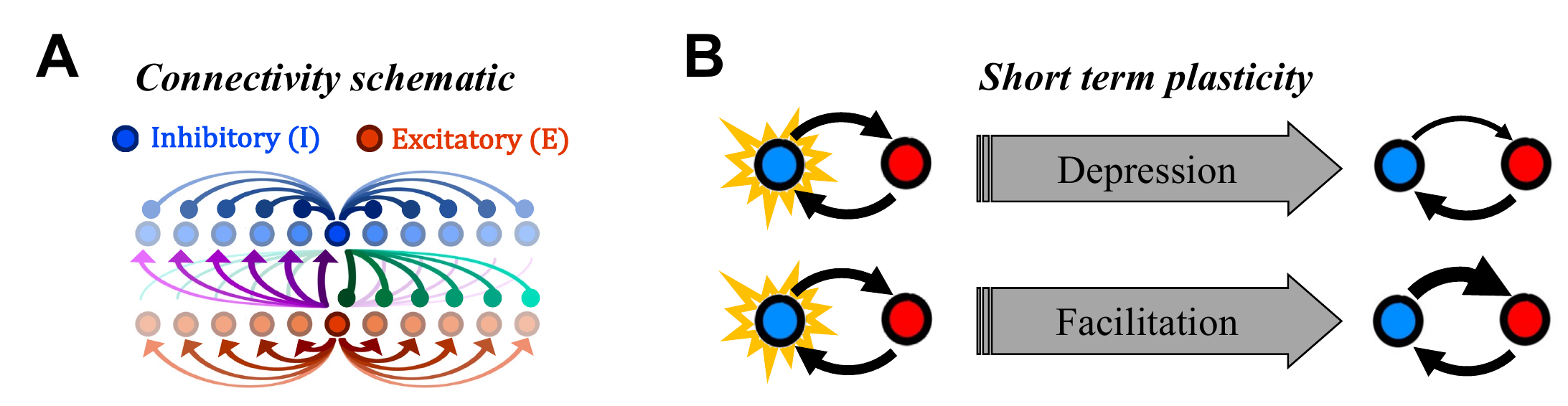}
  
  \caption{\textbf{Activity dependent pre-synaptic short term plasticity on separate excitatory (E) and inhibitory (I) neural sub-populations.} (A) Model schematic with separate E and I populations and the synaptic connections between them. (B) Schematic of the effects of activity dependent pre-synaptic short term plasticity on connection strength between neural populations. Top (Bottom): depression (facilitation) causes a decrease (increase) in the strength of synaptic connections originating from active populations, denoted by the thinning (thickening) arrows.}
  \label{fig:1}
\end{figure}
To analyze the effect of short term presynaptic plasticity (STP) (acting on the order of a few seconds) on separately evolving excitatory (E) and inhibitory (I) neural subpopulation activity (acting on the order of 10ms), we consider a set of stochastic (integro-differential) neural field equations with auxiliary variables evolving on slower timescales that modulate connectivity~\cite{WilsonCowan73,Blomquist05,Pinto2001b,cihak2022distinct}. Mean field equations for STP variables model the depletion of synaptic resources due to use (depression) and increase in the efficacy of release events (facilitation), for instance due to calcium signaling~\cite{tsodyks1998neural,jackman2017mechanisms}. Ultimately, the inclusion of this activity-dependent modulation of network architecture introduces additional slow timescales, which can alter the stability and dynamics of spatially structured solutions to the neural field.

Recurrent connectivity targeting neurons at position $x$ from $y$ at time $t$ is described the synaptic kernel $w(x-y)$, integrated against a nonlinearly filtered and STP-rescaled version of the synaptic input
(\Cref{fig:1}A). Synapses originating from recently activated neurons facilitate (strengthen) or depress (weaken) depending on plasticity parameterization (\Cref{fig:1}B). The dynamics of these synaptic variables is governed by up to four additional differential equations describing facilitation and depression on either the E or I neural subpopulations.
We include spatially structured noise in the E/I populations to account for neural activity fluctuations~\cite{faisal2008noise,Bressloff2010}.
\begin{subequations}
\begin{align}
    \tau_e du_e(x,t)=&\left[-u_e(x,t)+\mathcal{W}_{ee}(x,t)-\mathcal{W}_{ei}(x,t)\right]dt+\epsilon^{\frac{1}{2}} dW_{e}\\
    \tau_i du_i(x,t)=&\left[-u_i(x,t)+\mathcal{W}_{ie}(x,t)-\mathcal{W}_{ii}(x,t)\right]dt+\epsilon^{\frac{1}{2}}dW_{i}\\
    \tau_{qe}\partial_t q_e(x,t)=&-q_e(x,t)+\beta_{e}(q_{e0}-q_e(x,t))f_e(u_e(x,t))\\
    \tau_{qi}\partial_t q_i(x,t)=&-q_i(x,t)+\beta_{i}(q_{i0}-q_i(x,t))f_i(u_i(x,t))\\
    \tau_{re}\partial_t r_e(x,t)=&1-r_e(x,t)-\alpha_er_e(x,t)(1+q_e(x,t))f_e(u_e(x,t))\\
    \tau_{ri}\partial_t r_i(x,t)=&1-r_i(x,t)-\alpha_ir_i(x,t)(1+q_i(x,t))f_i(u_i(x,t)).
\end{align}\label{eq:model}
\end{subequations}
where $u_e(x,t)$ and $u_i(x,t)$ are the E and I synaptic input profiles at location $x$ at time $t$. We set $\tau_{e,i}=1$ and take time units to be 10ms, on the order of typical membrane and synaptic time constants~\cite{hausser1997estimating}. Recurrent connectivity terms are time dependent, shaped by facilitation $q_m$ and depression $r_m$ ($m=e,i$) dynamics
\begin{equation}
    \mathcal{W}_{mn}(x,t)=w_{mn}(x)*[r_n(x,t)(q_n(x,t)+1)f_n(u_n(x,t))]\hspace{1cm} m,n\in \{e,i\}.
\end{equation}
Facilitation dynamically strengthens synapses, due to activity-induced signaling, at a rate, $\beta_{e,i}$, up to a bound on available resources $q_{e0,i0}$. Depression depletes synaptic vesicle availability, making connections weaker at a rate $\alpha_{e,i}$ (See \Cref{fig:1}B). Rescaling the excitatory time units so $u_e$ evolves on the scale of 10ms and considering STP evolves on the timescale of seconds~\cite{tsodyks1998neural,jackman2017mechanisms} implies $\tau_q$ and $\tau_r$ are $\mathcal{O}(10^2)$. Chosen parameter ranges are given in \cref{tab:params}.

Sigmoidal firing rate functions, $f_n(u)=\frac{1}{1+e^{-\eta_n (u-\theta_n)}}$ with threshold $\theta_n$ and gain $\eta_n$, determine how synaptic input to neural population $n \in \{e,i\}$ is transferred to a normalized firing rate output.  Analytical results can be obtained in the high gain limit ($\eta_n \to \infty$), corresponding to a Heaviside firing rate function~\cite{Amari77,coombes2004evans}:
\begin{align}
    f_n(u) = H(u-\theta_n)=\begin{cases}1 & u-\theta_n \geq 0,\\ 0 & u-\theta_n < 0.\end{cases}\label{eq:2.2}
\end{align}
Solution dynamics can thus be tracked using {\em interface} equations associated with the evolution of level sets $u_e = \theta_e$ and $u_i = \theta_i$ in space and time~\cite{Coombes2012InterfaceJNeuro,Faye2018}.

Synaptic strength projecting from neurons at location $y$ to $x$ is described by the distance-dependent weight functions $w_{mn}(x-y)$ ($m,n \in \{ e, i\}$), taken to be symmetric exponentials for explicit calculations,   
\begin{align}
    w_{mn}(x)=A_{mn}e^{-\frac{|x|}{\sigma_{mn}}}, \hspace{10mm}m,n\in\{e,i\}
    \label{expwt}
\end{align}
where ranges for the amplitude and spatial scale $A_{mn},\sigma_{mn}\in \mathbb{R}^{\geq 0}$ (non-negative constants) are presented in \cref{tab:params}. E to E parameters are fixed (setting $A_{ee} =0.5$ and $\sigma_{ee} = 1$) to nondimensionalize based on $w_{ee}$. Other profiles are weaker and broader and determined by commonly observed $80\%$ E and $20\%$ I neuron ratio~\cite{abeles1991corticonics}.

\begin{table}[t!]
\label{tableparam}
{\footnotesize
  \caption{\textbf{Model parameters for Eq.~\cref{eq:model}}}\label{tab:params}
\begin{center}
  \begin{tabular}{|c|c|c|} \hline
   \bf{Parameter} & \bf{Definition}& \bf{Value}\\ \hline
   $A_{ee}$ &E-E strength &0.5\\ \hline $A_{ei},A_{ie}$ & I-E strength & 0.15\\ \hline
   $A_{ii}$ & I-I strength & 0.01 \\ \hline
   $\sigma_{ee}$ & E-E spatial scale & 1\\ \hline
   $\sigma_{ei},\sigma_{ie},\sigma_{ii}$  & other spatial scales & 2\\ \hline
   $\tau_e$,$\tau_i$ & E and I time constant & 1,1 \\ \hline
   $\theta_e,\theta_i$& firing thresholds vary & [0,0.5] \\ \hline
   $\tau_{qe},\tau_{qi}$ & facilitation time constant & 250 \\ \hline
   $\beta_{qe},\beta_{qi}$ & rate of facilitation & 0.01 \\ \hline
   $q_{e0},q_{i0}$ & synaptic resources available & [0,2]\\ \hline
   $\tau_{re},\tau_{ri}$ & depression time constant & 150\\ \hline
   $\alpha_{e},\alpha_{i}$ & rate of depression & [0,0.1]\\ \hline
  \end{tabular}
\end{center}
}
\end{table}

\begin{figure}[t!]
  \centering
  \includegraphics[width=0.90\textwidth]{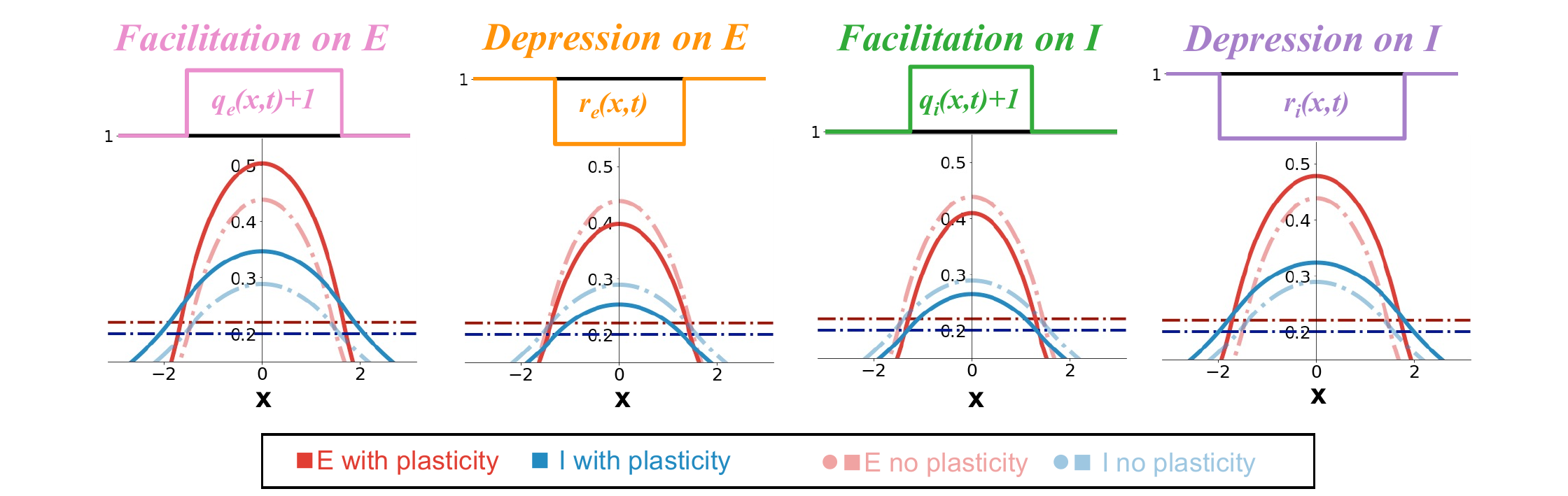}
  \vspace{-5mm}
  \caption{\textbf{Effects of short term plasticity variables on stationary solutions.} (Top) Facilitation/depression can increase/decrease synaptic efficacy within bumps. (Bottom) Facilitation on E (Depression on I) widens bumps and depression on E (Facilitation on I) narrows bumps. See \cref{sec:Deterministic} for detailed analysis.}
  \label{fig:2}
\end{figure}

Extending prior studies of deterministic E/I population models~\cite{Pinto2001b,Blomquist05,Compte2000} and those analyzing the impact of STP~\cite{kilpatrick2010stability,bressloff2011two}, we can explicitly construct bump (standing pulse) solutions and stability equations. Generally, recurrent excitation sustains both the E and I populations and inhibition prevents the spread of the E population activity. In the absence of noise, we obtain implicit formulas for the shape of even and translation symmetric bump and plasticity profiles and their dependence of STP parameters (See \cref{sec:Deterministic}, half-widths increase/decrease as in \cref{fig:2}).

Each bump has an {\em active region} over which neural activity ($u_e$ or $u_i$) is superthreshold (above $\theta_{e}$ or $\theta_{i}$) given by $[x_-^e(t),x_+^e(t)]$ and $[x_-^i(t),x_+^i(t)]$ and bounded by the interfaces (threshold crossings), 
$u_e(x_{\pm}^e(t),t)=\theta_e$ and $u_i(x_{\pm}^i(t),t)=\theta_i$~\cite{Amari77}. Threshold crossing conditions are constant in time for stationary bump half-widths, $a_e=({x}_+^e-{x}_-^e)/2$ and $a_i=({x}_+^i-{x}_-^i)/2$, but level set conditions become time-dependent when studying dynamics of perturbations. Active regions could be split into multiple disjoint segments for large perturbations, so we assume effects are weak enough to retain connected regions (See \cite{Coombes2012InterfaceJNeuro,Faye2018,Krishnan18} for elaborations on this problem). Stationary bump profiles are deformed by including effects of plasticity (\cref{fig:2}). Strengthening the effective excitation within bumps either by facilitating E synapses or depressing I synapses widens bumps. Complementarily, weakening effective excitation by depressing E or facilitating I synapses narrows bumps. Our subsequent stability and asymptotic analysis will study how both neural activity and synaptic efficacy variables respond to perturbation, and how their different timescales interact.

\begin{figure}[t!]
  \centering
  \includegraphics[width=0.85\textwidth]{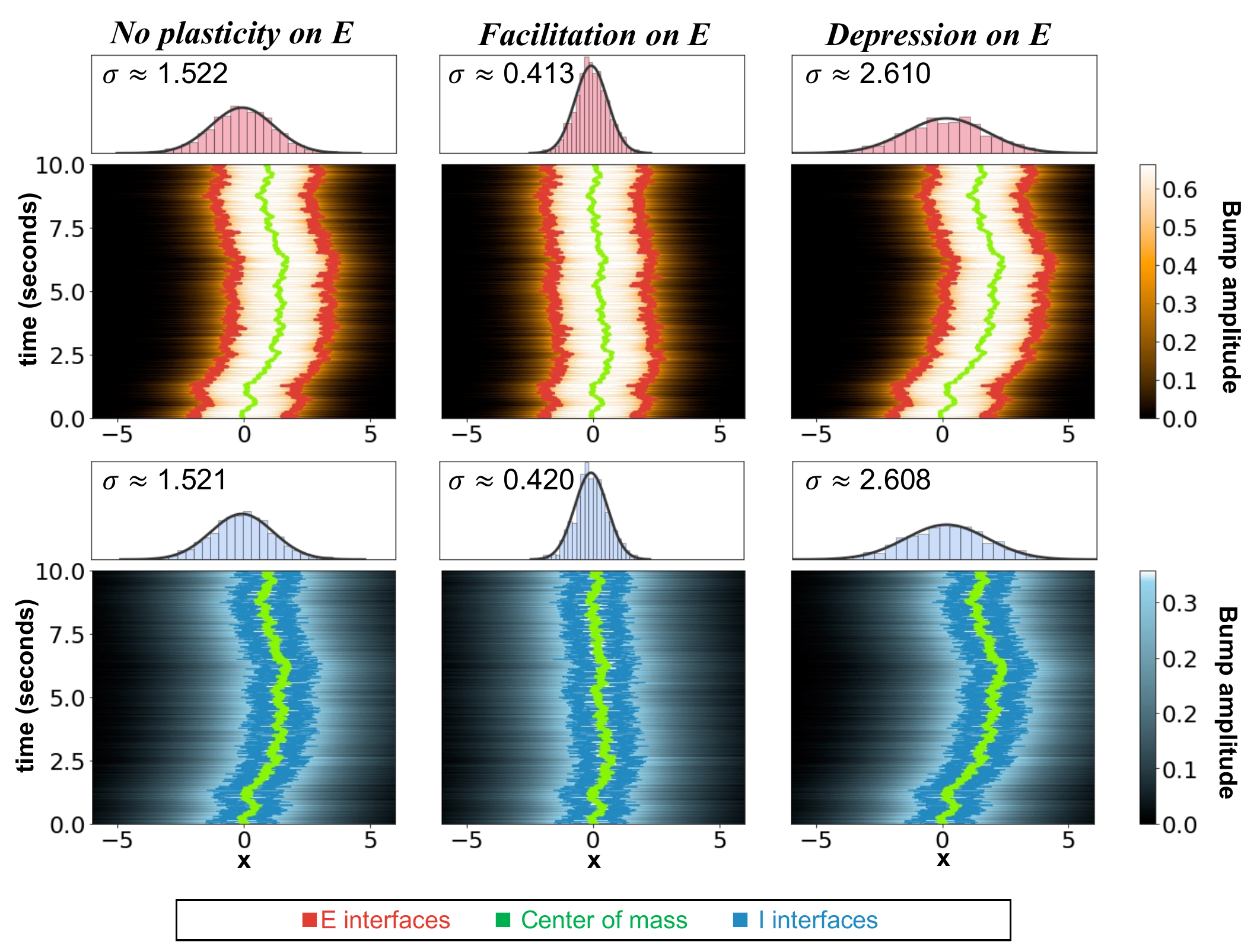} 
  \vspace{-4mm}
  \caption{\textbf{Wandering bumps and the effects of plasticity.} Numerical simulation of Eq.~(\ref{eq:model}), displaying excitatory and inhibitory bump activity over 10 seconds for the cases of: (Left) no STP with $q_{e0}=0$, $\alpha_e=0$, (Middle) facilitation on the E population with $q_{e0}=2$, $\alpha_e=0$, and (Right) depression on the E population with $q_{e0}=0$, $\alpha_e=0.006$. Centroids (green lines) denote the average position of E and I bump active regions as a function of time. Histograms show the final positions of centroids from 1000 simulations to more clearly demonstrate the effects of STP on bump wandering. These results show a decrease (increase) in wandering for facilitation (depression) on the E population, as seen in \cite{Kilpatrick18STP_FacSCIREP,Seeholzer19}. Other parameters: $A_{ii}=0$, $\theta_e=0.4$, $\theta_i=0.35$, $\epsilon=0.001$, $dx=0.005$ (space-step), and $dt=1$ms (timestep). Remaining parameters are as in \cref{tableparam}.}
  \label{fig:3}
\end{figure}

Stochastic motion of the bumps will be tracked by estimating the center of mass $ \Delta_{e}(t)$ ($ \Delta_{i}(t)$) of the active region of the E (I) bump
\begin{align}
    \Delta_{e}(t)= \frac{x_-^e(t)+x_+^e(t)}{2}, \hspace{2cm}
    \Delta_{i}(t)= \frac{x_-^i(t)+x_+^i(t)}{2}.\label{eq:centeromass}
\end{align}
The noise terms
$dW_{n}=\sqrt{|u_n(x,t)|}dZ_{n}(x,t)$ ($n\in\{e,i\}$) are  introduced with small amplitude $0 < \epsilon \ll 1$, which allows for asymptotic approximations of their effects. Increments of a spatially-extended Wiener processes have zero mean $\langle dZ_n(x,t)\rangle=0$ ($n \in \{e, i\}$) and spatial correlations $\langle dZ_n(x,t)dZ_n(y,s)\rangle=C_{n}(x-y)\delta(t-s)dtds$. 
Incorporating fluctuations causes bumps to `wander' due to their neutral stability to small perturbations~\cite{Laing2001,Kilpatrick13WandBumpSIAD}, along with profile deformation that tend to relax. We focus on how the polarity of STP on E and/or I bumps shapes the long term dynamics of their wandering (\cref{fig:3}). See \cref{app:num} for details on numerical methods.

We begin next section by analyzing bump stability, as a starting point for understanding their response to perturbations. Step discontinuities of the STP variables (\cref{fig:2}) require taking care in linearizing about solutions and treating singularities at bump interfaces.

\section{Deterministic Analysis}
\label{sec:Deterministic}

Stability calculations for bump solutions to Eq.~(\ref{eq:model}) in the absence of STP and noise are eased by using step nonlinear firing rate functions, localizing the problem to the threshold crossing locations~\cite{Amari77,Pinto2001b}. However, including plasticity then produces bump solutions with stepped STP variable profiles, complicating linear stability calculations. We will demonstrate how these singularities can be mollified and a piecewise linear description of stability can be formalized by a judicious change of variables~\cite{bressloff2011two,coombes2004evans}.

\subsection{Stationary Solutions}
\label{sec:stationary}

Assume solutions are stationary ($u_n(x,t)=U_n(x)$ with $n\in\{e,i\}$) and use the Heaviside firing rate function Eq.~\cref{eq:2.2}. We seek bumps with simply connected active regions ($U_n \geq \theta_n$) $[-a_n, a_n]$. Exploiting the solutions' translational invariance along $\mathbb{R}$ (opting for a center of mass at $x=0$) and expected evenness (i.e. $U_n(-x)=U_n(x)$) we obtain the system
\begin{subequations}
\begin{equation}
    \begin{split}
        U_n(x)=&{w_{ne}(x)*(1+Q_e(x))R_e(x)H(U_e(x)-\theta_e)}\\&-{w_{ni}(x)*(1+Q_i(x))R_i(x)H(U_i(x)-\theta_i)},
    \end{split}
\end{equation}
\begin{equation}
    R_n(x)=\begin{cases}\frac{1}{1+\alpha_n+\frac{q_{n0}\beta_{n}}{1+\beta_{n}}\alpha_n} & |x|\leq a_n\\
    1&|x| > a_n\end{cases}, \hspace{0.3cm}
    Q_n(x)=\begin{cases}
        \frac{q_{n0}\beta_{n}}{1+\beta_{n}} & |x|\leq a_n\\
        0 & |x|> a_n
    \end{cases}, \hspace{0.5cm} n\in\{e,i\},
\end{equation}\label{eq:3.1}
\end{subequations}
so neural activity bumps are continuous functions formed by convolving the weight kernels with scaled indicator functions for the active regions, and plasticity profiles are scaled indicator functions with jumps at the the bump edges. We find the explicit formulas for the bump half-widths, $a_e$ and $a_i$, by integrating the synaptic weight functions $w_{mn}(x)=A_{mn}e^{\frac{-|x|}{\sigma_{mn}}}$ ($m,n\in\{e,i\}$) over the active regions:
\begin{align}\int^a_{-a}w_{mn}(x-y)dy=\begin{cases}2A_{mn}\sigma_{mn}e^{\frac{-|x|}{\sigma_{mn}}}\sinh\left(\frac{a}{\sigma_{mn}}\right), & |x|>a,\\
2A_{mn}\sigma_{mn}\left[1-e^{\frac{-a}{\sigma_{mn}}}\cosh\left(\frac{x}{\sigma_{mn}}\right)\right], & |x|<a,
\end{cases}\label{eq:3.3}\end{align}

\noindent where $a \in \{ a_e, a_i\}$. Given that the terms $(1+Q_n(x))R_n(x)$ are constant within the intervals $[-a_n, a_n]$ ($n\in\{e,i\}$), substituting Eq.~\cref{eq:3.3} into Eq.~\cref{eq:3.1}, the threshold crossing conditions $\theta_n=U_n(\pm a_n)$, yield a piecewise and implicit set of equations for the half-widths, which depends on whether the E or I bump is wider, and assuming both E and I bumps have non-trivial active regions:
\begin{subequations}
\begin{align}
    \theta_e=&S_eA_{ee}\sigma_{ee}e^{\frac{-a_e}{\sigma_{ee}}}\sinh\left(\frac{a_e}{\sigma_{ee}}\right)-S_i\begin{cases}A_{ei}\sigma_{ei}e^{\frac{-a_e}{\sigma_{ei}}}\sinh\left(\frac{a_i}{\sigma_{ei}}\right), & a_e\geq a_i,\\
    A_{ei}\sigma_{ei}\left[1-e^{\frac{-a_i}{\sigma_{ei}}}\cosh\left(\frac{a_e}{\sigma_{ei}}\right)\right], & a_e<a_i,
    \end{cases}\\
    \theta_i=&-S_iA_{ii}\sigma_{ii}e^{\frac{-a_i}{\sigma_{ii}}}\sinh\left(\frac{a_i}{\sigma_{ii}}\right)+S_e\begin{cases}A_{ie}\sigma_{ie}e^{\frac{-a_i}{\sigma_{ie}}}\sinh\left(\frac{a_e}{\sigma_{ie}}\right), & a_i\geq a_e,\\
    A_{ie}\sigma_{ie}\left[1-e^{\frac{-a_e}{\sigma_{ie}}}\cosh\left(\frac{a_i}{\sigma_{ie}}\right)\right], & a_i<a_e.
    \end{cases}\\
    S_n=&2\frac{1+\beta_n+q_{n0}\beta_n}{(1+\beta_n)(1+\alpha_n)+q_{n0}\beta_n\alpha_n}, \hspace{2cm} n = e,i.
\end{align}\label{eq:3.4}
\end{subequations}

\begin{figure}[t!]
  \centering
  \includegraphics[width=0.85\textwidth]{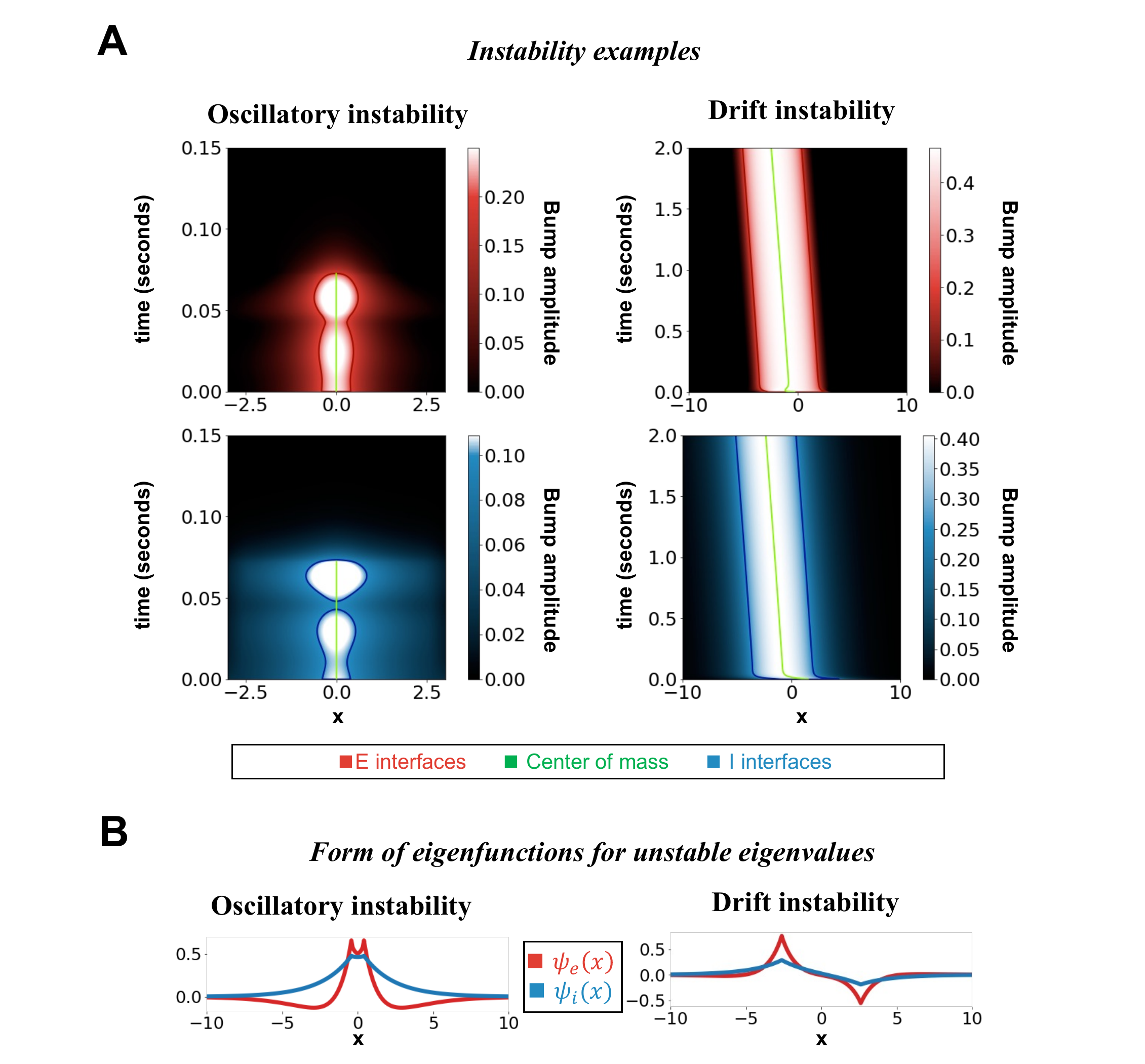}
  \caption{\textbf{Examples of destabilizing the broad bump.} (A) Left: `Oscillatory instabilities' leading to breathing of broad bumps with increasing amplitude resulting in bump collapse. An example of an oscillatory instability is shown here with parameters as in Table~\ref{tab:params} with $\theta_e=0.2$ and $\theta_i=0.1$, $q_{e0}=1$, $q_{i0}=1$, $\alpha_e=0.003$, $\alpha_i=0.1$, $dx=0.005$ and $dt=0.1$ms. Right: Sufficiently strong depression on E (facilitation on I) can generate a `drift instability,' leading to traveling waves. Parameters here are $\alpha_e=0.01$, $q_{i0}=2$ $\theta_e=0.2$, $\theta_i=0.25$ and the I population was initially perturbed left. Other parameters are as in \cref{tableparam}. (B) Examples of the eigenfunctions associated with unstable eigenvalues in the corresponding oscillatory (even) and drift (odd) instability cases.}  
  \label{fig:4}
\end{figure}

There is a second class of bump solutions where there is no active region in the I population (so that $U_i(x) < \theta_i$ for all $x \in \mathbb{R}$), each of which tends to be linearly unstable as there is no active inhibition to prevent the spread of excitation upon perturbation. Applying this assumption to Eq.~\cref{eq:3.1}, we obtain the simplified system 
with E bump half-width given by
\begin{equation}
    \theta_e=S_eA_{ee}\sigma_{ee}e^{\frac{-a_e}{\sigma_{ee}}}\sinh\left(\frac{a_e}{\sigma_{ee}}\right)
\end{equation}
with $U_i < \theta_i$ for all $x \in \mathbb{R}$, which for a unimodal bump occurs at $x=0$, so we require
\begin{align*}
    U_i(0) = \int_{-a_e}^{a_e} w_{ie} (y) dy = S_iA_{ie}\sigma_{ie}\left[1-e^{\frac{-a_e}{\sigma_{ie}}}\right] < \theta_i.
\end{align*}
Thus, one branch of `Broad' bumps has superthreshold active regions in both the E and I populations ($a_e>0$ and $a_i>0$). For sufficiently high thresholds, these solutions are often marginally stable solutions, but can destabilize into oscillations for sufficiently low firing rate thresholds (See \cref{fig:4}A and subsequent stability analysis). In contrast, when the I bump is subthreshold ($U_i(0)< \theta_i$), we obtain unstable `Narrow' solutions that function akin to a separatrix (in the space of active region widths) between the Broad solutions and the rest state. Narrow and Broad bumps annihilate in a discontinuous saddle-node bifurcation, due to the I Narrow bump grazing the threshold $\theta_i$ as in \cite{cihak2022distinct}. Including STP leads to solutions that can destabilize via a drift instability. We now derive stability results in detail.


\subsection{Eigenvalues and Noiseless Perturbations}
\label{sec:StabilityAnalysis}
Ignoring noise in Eq.~(\ref{eq:model}) and analyzing local stability of bumps to small perturbations, we first linearize in the sense of distributions as in \cite{Pinto2001b,Blomquist05,cihak2022distinct}. To treat jumps (and singularities) in plasticity profiles, we extend methods developed in \cite{kilpatrick2010stability,bressloff2011two}, performing a change of variables via integration to mollify bare delta distributions obtained from differentiating jumps. The result is a piecewise defined linear stability problem so that the response of bumps to perturbations depends on the direction they move the bump edges.

To begin our analysis of the evolution of small perturbations, we let $u_n(x,t)=U_n(x)+\epsilon \Psi_n(x,t)$,  $r_n(x,t)=R_n(x)+\epsilon \Theta_n(x,t)$, and
$q_n(x,t)=Q_n(x)+\epsilon \Lambda_n(x,t)$ ($n \in \{e,i\}$).
Substituting into Eq.~\cref{eq:model} (without noise), applying the stationary solutions, and simplifying yields: 
\begin{subequations}
\begin{equation}
    \begin{split}
        \tau_n\frac{\partial\Psi_n}{\partial t} + \Psi_n =&\sum_{m=e,i} {\mc S}^{m} \left( w_{nm}*\left[ (\Lambda_mR_m+\Theta_m (1+Q_m))H(U_m+\epsilon \Psi_m-\theta_m) \right] \right. \\
    &\left. +\frac{1}{\epsilon}w_{nm}*\left[ R_m (1+Q_m)[H(U_m+\epsilon \Psi_m-\theta_m)-H(U_m-\theta_m)] \right] \right)
    \end{split}
\end{equation}
\begin{equation} \label{eq:pdthet}
\begin{split}
    \tau_{rn}\frac{\partial\Theta_n}{\partial t}=&{-\Theta_n}-(\Lambda_nR_n+\Theta_n (1+Q_n))\alpha_{n}H(U_n+\epsilon \Psi_n-\theta_n)\\
    &-\frac{1}{\epsilon}\alpha_{n}R_n (1+Q_n)[H(U_n+\epsilon \Psi_n-\theta_n)-H(U_n-\theta_n)]
\end{split}
\end{equation}
\begin{equation} \label{eq:pdlamb}
\begin{split}
    \tau_{qn}\frac{\partial\Lambda_n}{\partial t}=&-\Lambda_n-\Lambda_n\beta_{n}H(U_n+\epsilon \Psi_n)\\
    &+\frac{1}{\epsilon}\beta_{n}(q_{n0}-Q_n)[H(U_n+\epsilon \Psi_n-\theta_n)-H(U_n-\theta_n)]
\end{split}
\end{equation}
\end{subequations}

\noindent where $n\in\{e,i\}$ and ${\mc S}^{e} = 1$ (${\mc S}^i = -1$) is the polarity of E (I) connectivity.
Perturbations to the bumps $\Psi_n$ can shift bump boundaries, described by the threshold crossing points, as $\pm a_n+\epsilon\kappa_n^{\pm}$ so that 
\begin{equation}
    u_n(\pm a_n+\epsilon\kappa_n^{\pm},t)=\theta_n, \label{eq:pert_thresh}
\end{equation}
\noindent where applying self-consistency to a linearization of Eq.~(\ref{eq:pert_thresh}) implies
\begin{align*}
    U_n(\pm a_n) + \epsilon\kappa_n^{\pm} U_n'(\pm a_n) + \epsilon\Psi (\pm a_n,t) \approx \theta_n \ \  \Rightarrow \ \  \kappa_n^{\pm} \approx\pm \frac{\Psi_n(\pm a_n,t)}{|U_n'(a_n)|}.
\end{align*}
Unperturbed STP profiles have jump discontinuities at bump edges, so the perturbations $\Theta_n$ and $\Lambda_n$ could undergo $\mathcal{O}(1/\ep)$ changes. However, if the spatial neighborhood is sufficiently small (${\mc O}(\ep)$), we can {\em shield} our linearization from these large changes by mollifying with a change of variables, introducing \textit{auxiliary fields} of the form~\cite{kilpatrick2010stability,bressloff2011two}: 
\begin{align*}
        M_{mn}(x,t)&=w_{mn}*\left[ \Lambda_n H(U_n+\epsilon \Psi_n-\theta_n) \right]=\int^{a_n+\epsilon\kappa^+_n(t)}_{-a_n+\epsilon\kappa^-_b(t)}w_{mn}(x-y)\Lambda_n(y,t)dy, \\
        N_{mn}(x,t)&=w_{mn}*\left[ \Theta_n  H(U_n +\epsilon \Psi_n -\theta_n) \right]=\int^{a_n+\epsilon\kappa^+_n(t)}_{-a_n+\epsilon\kappa^-_n(t)}w_{mn}(x-y)\Theta_n(y,t)dy,
\end{align*}
for $m,n \in \{e,i\}$, which will remain $\mathcal{O}(1)$ when $\Theta_n$ and $\Lambda_n$ are $\mathcal{O}(1/\epsilon)$ over an ${\mc O}(\ep)$ neighborhood. These auxiliary fields incorporate all the possible origin and target neural populations, expanding the set of linearized equations from six to ten. Differentiating with respect to $t$ yields
\begin{align*}
    \frac{\partial M_{mn}}{\partial t}&=w_{mn}* \left[ \frac{\partial \Lambda_n}{\partial t} \cdot H(U_n+\epsilon\Psi_n-\theta_n) \right] + {\mc O}(\ep), \\
    \frac{\partial N_{mn}}{\partial t}&=w_{mn}* \left[ \frac{\partial \Theta_n}{\partial t} \cdot H(U_n+\epsilon\Psi_n-\theta_n) \right] + {\mc O} (\ep).
\end{align*}
\noindent
Note, our leading order approximations must respect the sign of the perturbation at the bump edges determined by the sign of $\kappa_n^{\pm}$, resolved by piecewise-defining the limiting integrals~\cite{kilpatrick2010stability,bressloff2011two}:
\begin{align*}
     &\int^{a_n+\epsilon\kappa_n^+(t)}_{a_n}w_{mn}(x-y)R_n(y)(1+Q_n(y))dy\\ &\approx \epsilon\kappa_n^+\lim_{\epsilon\rightarrow 0^+}w_{mn}(x-a_n-\epsilon\kappa_n^+(t))R_n(a_n+\epsilon\kappa_n^+(t))(1+Q_n(a_n+\epsilon\kappa_n^+(t)))\\&=\epsilon\kappa_n^+(t)w_{mn}(x-a_n)G_n(\kappa_n^+(t)), \\
     &\int_{-a_n+\epsilon\kappa_n^-(t)}^{-a_n}w_{mn}(x-y)R_n(y)(1+Q_n(y))dy\\ &\approx -\epsilon\kappa_n^-\lim_{\epsilon\rightarrow 0^+}w_{mn}(x+a_n-\epsilon\kappa_n^-(t))R_n(-a_n+\epsilon\kappa_n^-(t))(1+Q_n(-a_n+\epsilon\kappa_n^-(t)))\\&=-\epsilon\kappa_n^-(t)w(x+a_n)G_n(-\kappa_n^-(t))
\end{align*}
where we define the piecewise constant function to takes the sign of the associated bump edge perturbation as its argument
\begin{align*}
    G_n(\kappa)=\begin{cases}
    1, & \kappa >0,\\
    \frac{1+\beta_{n}(1+q_{n0})}{(1+\alpha_n)(1+\beta_n)+q_{n0}\beta_n\alpha_n}, & \kappa <0.
    \end{cases}
\end{align*}
Thus, when a bump edge is perturbed outward ($\pm \kappa_n^{\pm} > 0$), the local dynamics is determined by a synaptic efficacy undisturbed by plasticity ($G_n =1$) whereas when a bump edge is perturbed inward ($ \pm \kappa_n^{\pm} < 0$), the local dynamics is determined by the plasticity-affected region.  Substituting in Eq.~(\ref{eq:pdthet}) and (\ref{eq:pdlamb}) for $\frac{\partial \Theta}{\partial t}$ and $\frac{\partial \Lambda}{\partial t}$, expanding in powers of $\epsilon$, collecting $\mathcal{O}(1)$ terms, approximating integrals, and simplifying yields a system of ten equations which we write in compact notation:
\begin{subequations} \label{eq:tdepstab}
\begin{equation}
    \begin{split}
        \tau_n\frac{\partial\Psi_n}{\partial t}+ \Psi_n=&\sum_{m=e,i} {\mc S}^m \left( \frac{(1+\beta_m)M_{nm}}{(1+\alpha_m)(1+\beta_m)+q_{m0}\beta_m\alpha_m}+\frac{1+\beta_m(1+q_{m0})}{1+\beta_m}N_{nm} \right. \\
        &\left. +\kappa_m^+ \cdot w_{nm}(x-a_{um})G_m(\kappa_m^+)-\kappa_m^- \cdot w_{nm}(x+a_{um})G_m(-\kappa_m^-) \right)
    \end{split}
\end{equation}
\begin{equation}
    \begin{split}
        \tau_{qn}\frac{\partial M_{mn}}{\partial t}=&-(1+\beta_{n})M_{mn}\\
    &+\beta_{n}q_{n0}[\kappa_n^+ \cdot w_{mn}(x-a_n)H(\kappa_n^+)-\kappa_n^- \cdot w_{mn}(x+a_n)H(-\kappa_n^-)],
    \end{split}
\end{equation}
\begin{equation}
    \begin{split}
        \tau_{rn}\frac{\partial N_{mn}}{\partial t}=&-\left(\frac{1+\beta_n+\alpha_n+\beta_n\alpha_n(1+q_{n0})}{1+\beta_n}\right)N_{mn}\\
    &-\alpha_n[\kappa_n^+ \cdot w_{mn}(x-a_n)H(\kappa_n^+)-\kappa_n^- \cdot w_{mn}(x+a_n)H(-\kappa_n^-)],
    \end{split}
\end{equation}
\end{subequations}
for $m,n\in\{e,i\}$. Stability of bumps can be determined by the rate of growth of separable solutions, determined by the exponential constant in $e^{\lambda t}(\Psi_n(x),M_{mn}(x),N_{mn}(x))$ for $m,n\in\{e,i\}$. Different classes of perturbations are determined according to their constituent functions' signs at the bump edges, each generating distinct linear operators under the requirement that perturbations do not change sign for $t>0$. Since the step nonlinearity localizes the stability problem to the bump edges~\cite{coombes2004evans}, we can restrict our attention to the function values and associated growth rates there, $x=\pm a_e$ and $x=\pm a_i$, determining eigenvalues/vectors for all (sixteen) cases of $\{{\rm sign}[\Psi_e(a_e)],{\rm sign}[\Psi_e(-a_e)],{\rm sign}[\Psi_i(a_i)],{\rm sign}[\Psi_i(-a_i)]\}$.

Our analysis confirms narrow bumps are unstable. A saddle node bifurcation occurs where the Broad and Narrow solutions annihilate one another. Plasticity can also destabilize broad bumps via {\em drift} bifurcations, generating traveling waves (\Cref{fig:4} right column). Drift instabilities are associated with real unstable eigenvalues.  Broad bumps can also destabilize via oscillatory perturbations, resembling the Hopf bifurcation of pure linear stability problems, which occurs as firing thresholds are reduced (\Cref{fig:4} left column). Note, such oscillatory perturbations break our assumptions on the fixed sign of perturbations throughout their evolution. However, we speculate that as long as a complex eigenvalue with positive real part is associated with each bump edge perturbation sign combination, then the instability will persist past each crossing. See \cref{app:unst} and \Cref{fig:unstableclass} for more details on this approach.

\begin{figure}[t!]
  \centering
  \vspace{-2mm}
  \includegraphics[width=0.85\textwidth]{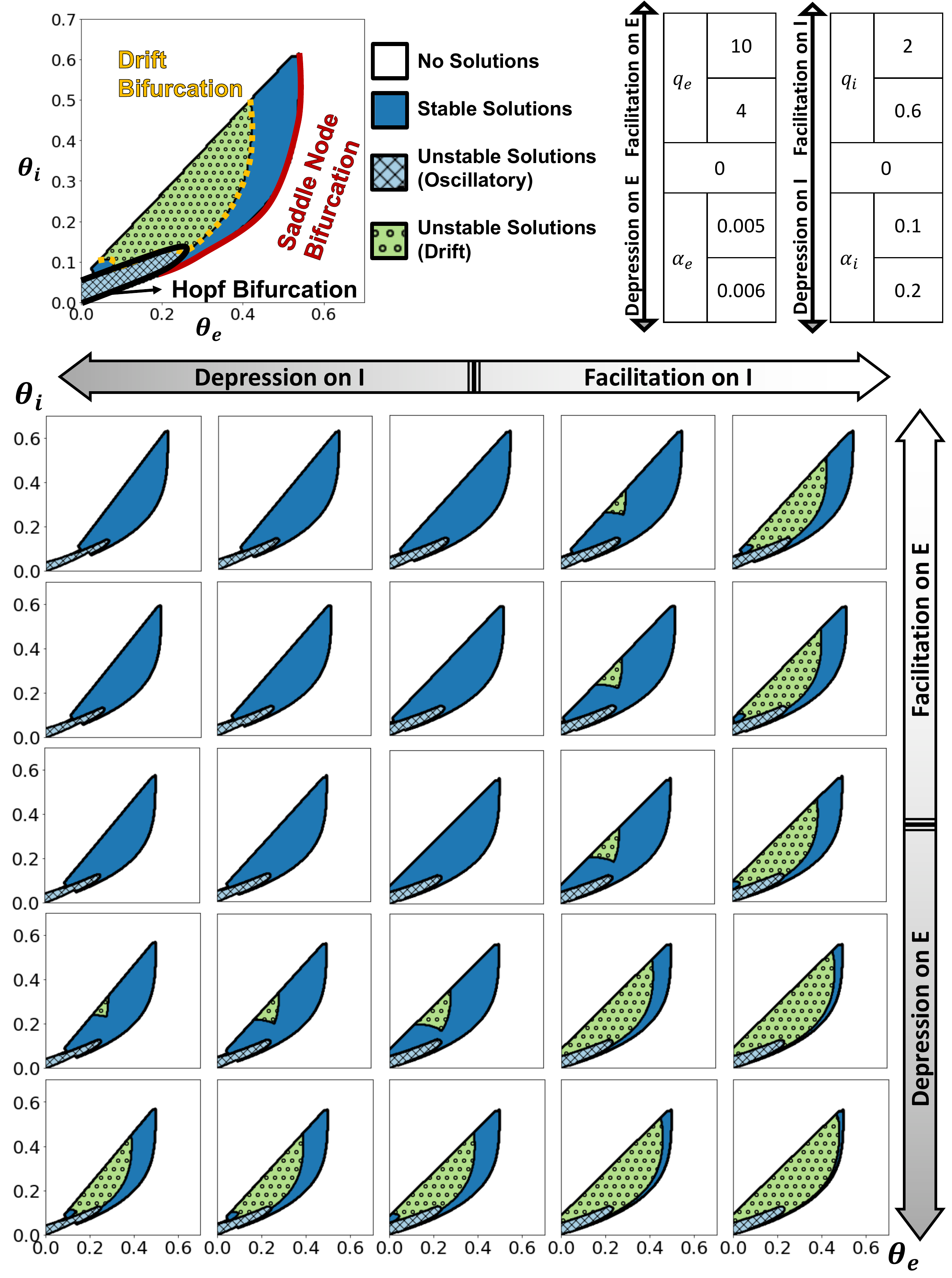}
  \vspace{-4mm}
  \caption{\textbf{Effects of short term plasticity variables on stability of bump solutions.} Parametric maps of stability as a function of firing thresholds $\theta_{e,i}$, determined via the largest eigenvalue and (for instabilities) the corresponding eigenfunction. See legend for color coding of stability/instability. Depression on E (facilitation on I) shrinks bumps and leads to an expansion of unstable regions in solution space. The widening of bumps' active regions due to facilitation on E (Depression on I) stabilizes the solution and tends to shrink the unstable regions. Additionally, we observe a slight expansion of the region where solutions exist. Plasticity parameters are varied along the grid of panels as shown in the small tables, all other parameters are as in \cref{tab:params}}
  \label{fig:5}
\end{figure}

To examine how the plasticity parameters affect the stability of solutions, we varied the strength of facilitation and depression of the E and I populations along with firing rate thresholds while fixing other parameters (\cref{fig:5}).
We found that as Facilitation on I or Depression on E increased, the drift instabilities arose and regions in which they were defined expanded. Alternatively increases in Facilitation on E or Depression on I reduced the size of the region in which drift instabilities occurred and could shift the region of bump solution existence.

To summarize, in this section we examined the local stability of bump solutions to small perturbations. Given the discontinuous nature of plasticity profiles, via our approach, we obtained piecewise smooth solutions for determining the stability of solutions. In the process we discovered the same discontinuities as in \cite{cihak2022distinct} in addition to a new type of discontinuity we call the drift instability, which behaves like a travelling wave causing the bumps to continuously drift in one direction dependent on the perturbation. Furthermore, with the addition of plasticity we found that solution existent regions could shift or destabilize due to changes in plasticity variables. 
\section{Low dimensional descriptions of stochastic bump motion}
\label{sec:Stochastic}
The stochastic wandering of bumps is a well validated mechanistic model associated with subject recall errors in delayed estimation tasks \cite{Compte2000,Wimmer14,panichello2019error}. Incorporating dynamic fluctuations into the neural field model Eq.~(\ref{eq:model}) (taking $\epsilon>0$) causes bumps to wander as a Brownian particle due to their marginal stability. Moving beyond the stabilizing effect of short term facilitation on a single neural population model observed in \cite{Kilpatrick18STP_FacSCIREP}, here we discriminate effects of short term plasticity as they depend on the polarity of synapses and their adaptive dynamics. Low-dimensional descriptions of stochastic bump motion are derived by exploiting the separation of timescales between neural activity and plasticity variables considering both a {\em strongly coupled limit} (which assumes the E and I bump remain close) and an {\em interface method} (which tracks distinct movements of bump edges). In our current work, we build on our analysis of the network with no plasticity~\cite{cihak2022distinct} by exploiting the slow timescale of plasticity variables to develop weak asymptotic approximations that are valid on both short and long timescales.

Our asymptotic equations are derived by weakly perturbing the position of stationary bumps ($U_n(x-\Delta_n(t))$ with $n\in\{e,i\}$) with the addition of small and slow plasticity variables, $q_n(x,t)$ and $r_n(x,t)$ and weak multiplicative noise, $W_n(x,t)$ ($|q_n|\ll 1$, $|r_n-1|\ll 1$, $|dW_n|\ll 1$, $\tau_{qn},\tau_{rn} \gg \tau_n$ for $n\in\{e,i\}$). Neural activity $u_n(x,t)$ will rapidly equilibrate to a quasi-steady-state determined by the slowly evolving plasticity profiles. Thus we have
\begin{align*}
    u_n(x,t)&={U}_n(x-\Delta_n(t))+\psi_n(x-\Delta_n(t),t)+\dots,\hspace{5mm}n\in\{e,i\}
\end{align*}
where $|\Delta_n(t)|\ll 1$ (with $|d\Delta_n|\ll1$) is the perturbation of the position of the bump with stationary profile $U_n$, and $\psi_n$ are profile perturbations with $|\psi_n|\ll1$. Before deriving the full asymptotic equations for stochastic bump motion, we will develop a low-dimensional model for the slowly evolving plasticity profiles.

\subsection{Plasticity profile evolution}
\label{s:plastprof}
Given $\Delta_n(t)$ is the centroid of bump $n \in \{ e, i\}$ at time $t$,  we expect that $q_n(x,t)$ ($r_n(x,t)$) will resemble $Q_n(x-\Delta_{qn})$ ($R_n(x-\Delta_{rn})$) where $\Delta_{qn}$ ($\Delta_{rn}$) is the slow evolution of the centroids of $q_n$ ($r_n$) for $n\in\{e,i\}$. 
To estimate each plasticity profile centroid we take the weighted average of $q_n$ or $1-r_n$, $n \in \{ e, i \}$.
To model the evolution of $\Delta_{qn}(t)$ and $\Delta_{rn}(t)$ we linearize the system about $u_n(x,t)=U_n(x-\Delta_n(t))$, so that $q_n(x,t)=Q_n(x-\Delta_n(t))+\Lambda_n(x,t)$, $r_n(x,t)=R_n(x-\Delta_n(t))+\Theta_n(x,t)$, finding for $f_n(u) = H(u-\theta_n)$ that
\begin{subequations}
\begin{align}
        -U_n'(x)d\Delta_n (t)&=\left[{w_{ne}(x)*(R_e(x)+\Theta_e(x,t)-1+(Q_e(x)+\Lambda_e(x,t)))H(U_e(x)-\theta_e)}\right. \nonumber \\
   &  \hspace{-1.3cm} \left.{-w_{ni}(x)*(R_i(x)+\Theta_i(x,t)-1+(Q_i(x)+\Lambda_i(x,t)))H(U_i(x)-\theta_i)}\right]dt+dW_{n}, \\
    &\tau_{qn}\frac{\partial\Lambda_n}{\partial t}=-\Lambda_n-\Lambda_n\beta_{n}H(U_n-\theta_n), \\
    &\tau_{rn}\frac{\partial\Theta_n}{\partial t}={-\Theta_n}-(\Theta_n(1+Q_n))\alpha_nH(U_n-\theta_n).
\end{align}
\end{subequations}
To mollify the piecewise smooth dynamics emerging in the perturbative equations due to the step nonlinearity, eight auxiliary fields are constructed by convolving step and perturbation products with weight functions~\cite{bressloff2011two}, $M_{mn}(x,t)=w_{mn}(x)*\Lambda_n H(U_n-\theta_n)$ and four of the form $N_{mn}(x,t)=w_{mn}(x)*\Theta_n H(U_n-\theta_n)$ whose evolution equations can be derived as
\begin{align*}
    \tau_{qn}\partial_t{M_{mn}} &=-(1+\beta_n)M_{mn}, \hspace{1cm}
        \tau_{rn}\partial_t{N_{mn}} =-\left(1+\frac{1+\beta_n(1+q_{n0})}{1+\beta_n}\alpha_n\right)N_{mn}.
\end{align*}

\begin{figure}[t!]
  \centering
  \includegraphics[width=0.9\textwidth]{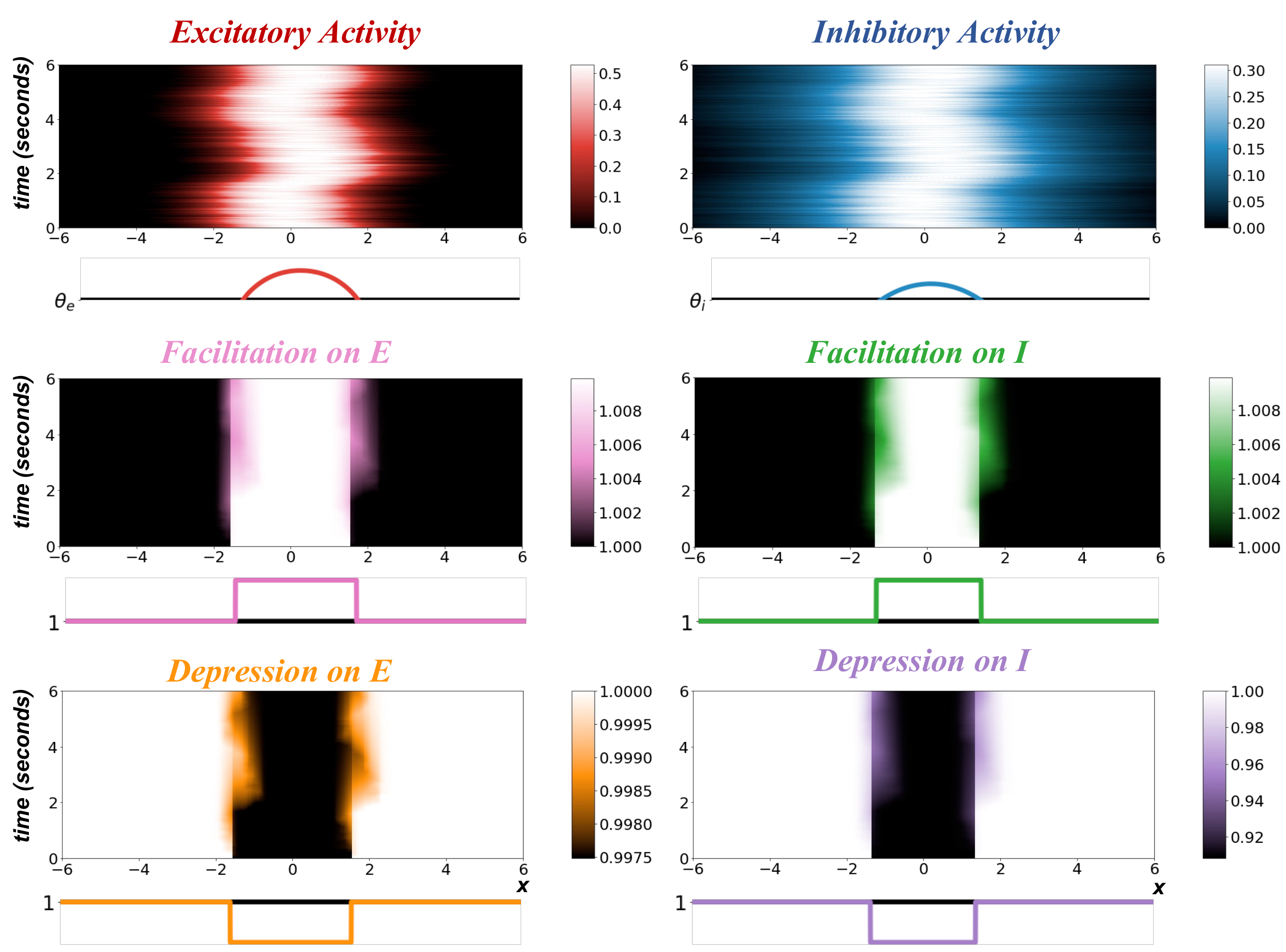}
  \caption{\textbf{Synaptic plasticity profile deformation in a single simulation} Heat maps represent the evolution of neural activity and synaptic plasticity profiles over time. Colorbars denote amplitudes. Below each heat map is the initial profile.
  Observe that the slower reacting synaptic plasticity profiles wander less and experience profile deformation, namely smoothing at the edges. Parameters are as in Table~\ref{tab:params} with $\theta_e=0.3$ and $\theta_i=0.25$, $q_{e0}=1$, $q_{i0}=1$, $\alpha_e=0.003$, $\alpha_i=0.1$, $dx=0.005$ and $dt=0.1$ ms.}
  \label{fig:6}
\end{figure}

 Via an assumption of separation of variables ($M_{nm}(x,t)=\overline{M}_{nm}(x)e^{\lambda_a t}$ and \\$N_{nm}(x,t)=\overline{N}_{nm}(x)e^{\lambda_b t}$) we find that $\lambda_a=-(1+\beta_{qn})/\tau_{qn}$ and \\$\lambda_b=-\left(1+\frac{1+\beta_n(1+q_{n0})}{1+\beta_n}\alpha_n\right) /\tau_{rn}$, the rates of decay of the perturbations of plasticity centroids from the activity centers $\Delta_n(t)$. Hence, the plasticity variables' centroids, $\Delta_{qn}$ and $\Delta_{rn}$, evolve according to the equations
\begin{subequations}
\begin{align}
    \tau_{qn} d\Delta_{qn} &=-(1+\beta_{qn})(\Delta_{qn}-\Delta_n), \\
        \tau_{rn} d\Delta_{rn}& =-\left(1+\frac{1+\beta_n(1+q_{n0})}{1+\beta_n}\alpha_n\right)(\Delta_{rn}-\Delta_n).
\end{align}
\end{subequations}
Furthermore, over time the stochastic motion of neural activity bumps $u_n(x,t)$ will smooth the initially sharp stepped edges of the plasticity profiles as shown in stochastic simulations (\cref{fig:6}).

This effect is similar to the blurring of sharp edges during image degradation (See \cref{app:blur} and \cite{smith1998characterization, ji2008motion, solomon2011fundamentals}). Plasticity profiles act as the {\em image} of an {\em object} (neural activity), blurring over time (See also \cref{fig:platypusBlur}),
which we can describe using an ansatz given by the convolution
\begin{equation}
    s(x,t)=\int_{- \infty}^{\infty} b(x-y, t)z(y)dy,
\end{equation}
where $s(x,t)$ is the blurred profile due to the evolution of the bump, $b(x,t)$ is the blurring kernel (or {\em point spread function}) and $z(x)$ is the input image (initial profile of the plasticity variable). The blurring kernel is assumed Gaussian, due to the Brownian motion typical of noise-driven bump wandering~\cite{Kilpatrick13WandBumpSIAD}, but the
standard deviation relaxes over long timescales, due to the slow kinetics of plasticity
\begin{align}
    \sigma (t)=\sqrt{\epsilon (|U_e(x)|+|U_i(x)|) \tau_{plast}}(1-e^{-t/\tau_{plast}}), \label{sigplast}
\end{align}
where $\tau_{plast}$ may be $\tau_{qn}$ or $\tau_{rn}$, $n \in \{ e, i\}$, so $\sigma(t) \to \sqrt{\epsilon (|U_e(x)|+|U_i(x)|) \tau_{plast}}$ as $t \to \infty$. Convolving the top hat-shaped initial profiles of $q_n(x,t)$ and $1-r_n(x,t)$ (the images) with the Gaussian blurring kernel, and including the center of mass shifts $\Delta_{qn}$ and $\Delta_{rn}$ we approximate evolution of the plasticity variables as
\begin{subequations} \label{eq:qrplastprof}
    \begin{align}
        q_n(x-\Delta_{qn},t) &\approx \frac{q_{n0}\beta_n}{1+\beta_n} \cdot E(x-\Delta_{qn};-a_n, a_n, \sigma (t)), \\
        1-r_n(x-\Delta_{rn},t) & \approx \frac{\alpha_n(1+\beta_n(1+q_{n0}))}{(1+\beta_n)(1+\alpha_n)+q_{n0}\beta_n\alpha_n} \cdot E(x-\Delta_{rn};- a_n, a_n, \sigma (t)),
    \end{align}
\end{subequations}
where profiles are now described using scaled error function~\cite{solomon2011fundamentals},
\begin{align*}
    E(x; x_1,x_2, \sigma) = \frac{1}{2}\left({\rm erf}\left( \frac{x_2 -x}{\sqrt{2\sigma^2}}\right)-{\rm erf}\left( \frac{x_1 - x}{\sqrt{2\sigma^2}}\right)\right).
\end{align*}

\begin{figure}[t!]
  \centering
  \includegraphics[width=\textwidth]{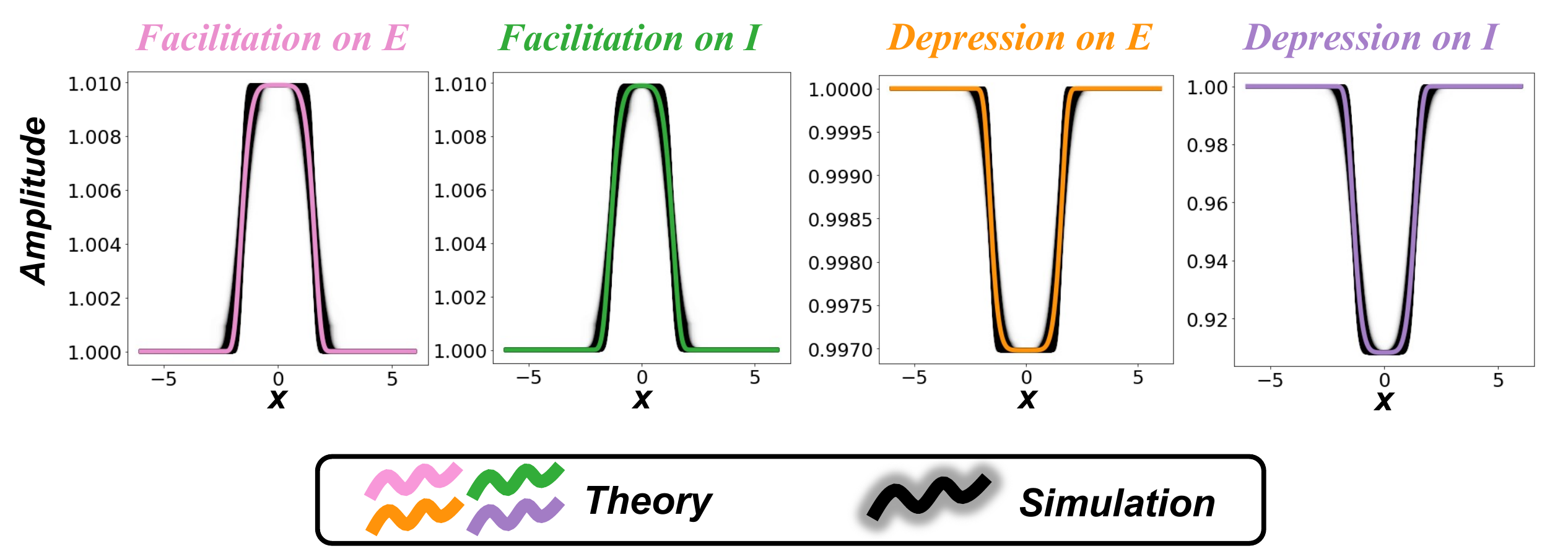}
  \vspace{-5mm}
  \caption{\textbf{Plasticity profile deformation theory compared with simulation.} Performing 10,000 Monte Carlo simulations of four seconds each, we compute the final profiles, recentered to zero and superimpose them with low transparency. Theoretical approximation for profile deformation matches fairly well at the average of these simulations. Parameters are as in Table~\ref{tab:params} with $\theta_e=0.3$ and $\theta_i=0.25$, $q_{e0}=1$, $q_{i0}=1$, $\alpha_e=0.003$, $\alpha_i=0.1$, $dx=0.005$ and $dt=0.1$}
  \label{fig:7}
\end{figure}

As time $t\rightarrow 0^+$, $\sigma (t) \to 0$, so $E(x;x_1, x_2, \sigma) \to \frac{1}{2} \left[ H(x_2 - x) - H(x_1 - x) \right]$, so we recover the initial profiles at $t=0$. Note, this theory also accurately captures the evolution and long term profile shape of the plasticity variables across many simulations (\cref{fig:7}).

\subsection{Strongly coupled limit approximation}
Our first approach to deriving stochastic evolution equations for bumps assumes profiles of E and I activity stay close as they move~\cite{cihak2022distinct}, the {\em strongly coupled limit},
$\Delta(t) \equiv \Delta_{e}(t)=\Delta_i(t)$, so that
\begin{align*}
    u_n(x,t)&={U}_n(x-\Delta(t))+\psi_n(x-\Delta(t),t)+\dots,\hspace{5mm}n\in\{e,i\}.
\end{align*}
Plugging in this ansatz into the neural activity equations in Eq.~(\ref{eq:model}) yields:
\begin{equation}
    \begin{split}
        -\tau_n U_n' d\Delta &+\tau_n d\psi_n =\left[-U_n -\psi_n \right.\\
    &+{w_{ne}*\left[ r_e(x+\Delta,t)(1+q_e(x+\Delta,t))H(U_e+\psi_e - \theta_e) \right] }\\
    &\left.-{w_{ni}*\left[ r_i(x+\Delta,t)(1+q_i(x+\Delta,t))H(U_i+\psi_i - \theta_i) \right]}\right]dt+ dW_{n},
    \end{split}
\end{equation}
for $n\in\{e,i\}$.
Cancelling the implicit equation for the stationary solution and collecting $\mathcal{O}(1)$ terms yields the system
\begin{equation}
    \begin{split}
        \tau_nd\psi_n+\psi_n
   &- w_{ne}* \left[ H'(U_e - \theta_e)\psi_e \right] dt + w_{ni}* \left[ H'(U_i - \theta_i) \psi_i \right] dt\\
   &=\tau_nU_n' d\Delta +\left[w_{ne}*(r_e(x+\Delta,t)-1+q_e(x+\Delta(t),t))H(U_e - \theta_e)\right.\\
   &\left.-w_{ni}*(r_i(x+\Delta,t)-1+q_i(x+\Delta(t),t))H(U_i - \theta_i)\right]dt+dW_{n}
    \end{split}
\end{equation}
with $n\in\{e,i\}$.
Thus the system describing bump profile perturbations evolution is
\begin{equation}
    \begin{split}
        &\begin{pmatrix}
    \tau_e d\psi_e\\ \tau_i d\psi_i
    \end{pmatrix}-\mathcal{L}\begin{pmatrix}
    \psi_e\\ \psi_i
    \end{pmatrix}dt=\begin{pmatrix}
    \tau_eU_e'\\\tau_i U_i'
    \end{pmatrix}d\Delta +\begin{pmatrix}
    dW_e\\dW_i
    \end{pmatrix} \\&+\begin{pmatrix}
    w_{ee}*(q_e(x+\Delta(t),t))H(U_e - \theta_e)-w_{ei}*(q_i(x+\Delta(t),t))H(U_i - \theta_i)\\
    w_{ie}*(q_e(x+\Delta(t),t))H(U_e - \theta_e)-w_{ii}*(q_i(x+\Delta(t),t))H(U_i- \theta_i)
    \end{pmatrix}dt\\&+\begin{pmatrix}
    w_{ee}*(r_e(x+\Delta(t),t)-1)H(U_e-\theta_e)-w_{ei}*(r_i(x+\Delta(t),t)-1)H(U_i-\theta_i)\\
    w_{ie}*(r_e(x+\Delta(t),t)-1)H(U_e-\theta_e)-w_{ii}*(r_i(x+\Delta(t),t)-1)H(U_i-\theta_i)
    \end{pmatrix}dt
    \end{split}
    \label{eq:coupledsystem}
\end{equation}
where the linear operator ${\mc L}$ is as described in \cite{cihak2022distinct}:
\begin{align*}
    \mathcal{L}\begin{pmatrix}
    r\\s
    \end{pmatrix}=\begin{pmatrix}
    -r+w_{ee}*[H'(U_e-\theta_e)r]- w_{ei}*[H'(U_i - \theta_i)s]\\
    -s+w_{ie}*[H'(U_e-\theta_e)r]- w_{ii}*[H'(U_i - \theta_i)s]
    \end{pmatrix},
\end{align*}
where we can define $H'(U_n(x)-\theta_n) = |U_n'(a_n)|^{-1} \sum_{y=\pm a_n} \delta (x-y)$, since
\begin{align*}
    \delta(x + a_n) - \delta(x - a_n) = \frac{d}{dx}H(U_n(x) - \theta_n) = U_n'(x) H'(U_n(x)-\theta_n).
\end{align*}
We enforce a bounded solution of Eq.~(\ref{eq:coupledsystem}) by requiring that the inhomogeneous part is orthogonal to the nullspace of the adjoint operator, 
\begin{align*}
    {\mc L}^* \begin{pmatrix}
    a \\b 
    \end{pmatrix} = \begin{pmatrix} -a + H'(U_e-\theta_e) \left[ w_{ee}*a+w_{ie}*b \right] \\ -b - H'(U_i-\theta_i) \left[ w_{ei}*a+w_{ii}*b \right] \end{pmatrix}
\end{align*}
The nullspace is spanned by a single vector~\cite{cihak2022distinct},
\begin{align}
    \begin{pmatrix}
    \phi_1(x)\\
    \phi_2(x)
    \end{pmatrix}
    =
    \begin{pmatrix}
    \delta(x+a_e)-\delta(x-a_e)\\
    \mathcal{B}[\delta(x+a_i)-\delta(x-a_i)]
    \end{pmatrix},
\end{align}
with $\mathcal{B}=-\frac{w_{ei}(a_e-a_i)-w_{ei}(a_e+a_i)}{w_{ie}(a_e-a_i)-w_{ie}(a_e+a_i)}$.
The inner product $\langle \mathbf{u}, \mathbf{v} \rangle = \int_{- \infty}^{\infty} \mathbf{v}^*(x) \mathbf{u}(x) d x$ of the nullspace with the inhomogeneous part (the right half) of Eq. (\ref{eq:coupledsystem}) is computed and rearranged to find
\begin{subequations}
\begin{equation}
    d\Delta=\mathcal{K}(\Delta,q_n,r_n,t)dt+\Sigma d\xi(t), \hspace{3mm} n\in\{e,i\},
\end{equation}\label{eq: coupledPred}
where
\begin{align*}
    \Sigma^2/2&=\epsilon \frac{\Gamma_e+\Gamma_0+\mathcal{B}^2\Gamma_i}
    {2\left[\tau_e|U_e'(a_e)|+\mathcal{B}\tau_i|U_i'(a_i)|\right]^2},\\
    \Gamma_n &=\theta_n[C_n(0)-C_n(2a_{n})], \ \ \ \Gamma_0=2\mathcal{B}\sqrt{\theta_e\theta_i}[C_c(a_e-a_i)-C_c(a_e+a_i)],
\end{align*}
and
\begin{align*}
\mathcal{K}=&-\frac{\sum_{n = e,i} {\mc S}^m \langle
    \rho_n(x), (q_n(x+\Delta(t),t) + r_n(x+\Delta,t)-1) H(U_n(x)-\theta_n) \rangle}{2\left[\tau_e|U_e'(a_e)|+\mathcal{B}\tau_i|U_i'(a_i)|\right]},\\
_n(x)=&[w_{en}(-a_e-x)-w_{en}(a_e-x) +\mathcal{B}(w_{in}(-a_i-x)-w_{in}(a_i-x))],
\end{align*}
\end{subequations}
Note, the plasticity profiles $q_n$ and $r_n$ may take any form as long as they are weak perturbations from 0 and 1 respectively. Subsequently, applying the plasticity profile approximation developed in \cref{s:plastprof}, we can precisely approximate how the plasticity profile deforms and shifts over time, allowing us to write a system of evolution equations of simply the activity and plasticity profiles center of mass: 
\begin{subequations}
\begin{align}
    d\Delta&=\mathcal{K}(\Delta,\Delta_{qe},\Delta_{qi},\Delta_{re},\Delta_{ri},t)dt+\Sigma d\xi(t), \\
    \tau_{qn} d\Delta_{qn}&=-(1+\beta_{qn})(\Delta_{qn}-\Delta) dt, \\
        \tau_{rn} d\Delta_{rn}&=-\left(1+\frac{1+\beta_n(1+q_{n0})}{1+\beta_n}\alpha_n\right)(\Delta_{rn}-\Delta) dt,
\end{align}
\end{subequations}
where $n \in \{e,i\}$ with $\mathcal{K}$ defined in \cref{eq: coupledPred} and $q_n$ and $r_n$ are as approximated in Eq.~(\ref{eq:qrplastprof}).

This approximation now only relies on the initial profiles and the evolution of the centers of mass of each profile, yielding a set of five differential equations over the six of the original system. 
Furthermore, this system is less computationally intensive and can be readily implemented numerically to approximate variances of the centers of mass. Effectively, bump centroids are assumed to be strongly coupled and so co-located, moving rapidly compared to the slowly evolving plasticity profiles, given only by initial profiles shifted by their centroid resulting in a closed and finite dimensional stochastic system of differential equations for the activity-plasticity evolution.

\subsection{Interface Based Approximation Theory}
The strong coupling approximation assumes E/I bumps remain close together, but of course, the independent of fluctuations in each population may cause activity profiles may stray from one another. We previously found simultaneous perturbations moving E and I bumps in the opposite direction tended to be momentarily {\em enhanced}, due to both bumps transiently straying from each other before settling at a new position~\cite{cihak2022distinct}. Such effects can only be described with a theory that tracks E and I bumps as separate but coupled.
Moving beyond standard linearization, {\em interface methods} can be used to account for interactions between bumps and other nonlinear effects arising in the transient response of bumps to perturbations. Originally applied to weakly perturbed bumps in neural fields with step nonlinearities~\cite{Amari77}, these methods focus on the evolution of dynamics at the edges of patterns, more recently used to track the phase of traveling front onset~\cite{Faye2018} or propagation in inhomogeneous neural fields~\cite{coombes2011pulsating} and complex (even labyrinthine) patterns in planar neural fields~\cite{Coombes2012InterfaceJNeuro}. Notably, it is possible to obtain fully consistent and exact representations of pattern dynamics by tracking edge positions and the gradient at those edges over time. Stochastic motion of multiple interacting bumps \cite{Krishnan18} or bumps E/I neural populations \cite{cihak2022distinct} can be approximated by a low-dimensional set of nonlinear Langevin equations by truncating the full theory. Here we further this theory incorporate effects of short term plasticity.

Interfaces enclose active regions of E/I bumps: $[x_-^e(t),x_+^e(t)]$ and $[x_-^i(t),x_+^i(t)]$ respectively.
Tracking stochastic dynamics of the level set conditions $u_e(x_{\pm}^e(t),t)=\theta_e$ and $u_i(x_{\pm}^i(t),t)=\theta_i$ allow us to determine effects of noise on bump positions over time. We differentiate and immediately approximate by dropping $\mathcal{O}(\epsilon)$ terms, yielding consistency equations for the interfaces $x_{\pm}^{e,i}(t)$:
    \begin{align}
\partial_xu_n(x_{\pm}^n(t),t)dx_{\pm}^n+du_n(x_{\pm}^n(t),t)=0, \hspace{1cm} n \in \{ e, i\}. \label{eq:consistency}
\end{align}
To first order, we assume the spatial gradients at the interfaces remain constant and odd symmetric throughout the evolution:
\begin{align*}
    {\mathcal G}_n&=|U_n'(a_n)|\approx \partial_x u_n(x_-^n(t),t)=-\partial_x u_n(x_+^n(t),t), \ \ \ \ n \in \{e,i\}.
\end{align*}
Substituting this approximation into 
Eq.~(\ref{eq:consistency}) and in the model Eq.~(\ref{eq:model}) with a step nonlinearity, and truncating, yields the following multiscale system of nonlinear Langevin equations describing the stochastic evolution of the interfaces subject to short term synaptic plasticity:
\begin{subequations}
\begin{align}
    \tau_n dx_{\pm}^n&=-\frac{1}{{\mathcal G}_n}\left(
    \left[-\theta_n+\Upsilon_{ne}(x_{\pm}^n; x_-^e, x_+^e)-\Upsilon_{ni}(x_{\pm}^n;x_-^i,x_+^i)
    \right]dt +\epsilon^{\frac{1}{2}}dW_e(x_{\pm}^n,t)\right), \\
    &\tau_{qn}\partial_t q_n(x,t)=-q_n(x,t)+\beta_{n}(q_{n0}-q_n(x,t))I_{[x_-^n(t),x_+^n(t)]}(x),\\
    &\tau_{rn}\partial_t r_n(x,t)=1-r_n(x,t)-\alpha_nr_n(x,t)(1+q_n(x,t))I_{[x_-^n(t),x_+^n(t)]}(x),
\end{align}\label{eq:interface}
\end{subequations}
where $n \in \{ e, i\}$ and the coupling between interfaces is given by integrals of the dynamically scaled weight functions over the active regions
\begin{align*}
    \Upsilon_{nm}(x;x_-^m, x_+^m) = \int^{x_+^m}_{x_-^m}r_m(y,t)(q_m(y,t)+1)w_{nm}(x-y)dy,
\end{align*}
with $m,n \in \{ e, i \}$ and we define the indicator function $I_A(x)=1$ if $x \in A$ and $0$ otherwise.
Interface equations are collapsed to describe the evolution of bump and plasticity profile centroids by applying the definitions Eq.~(\ref{eq:centeromass}) to Eq.~(\ref{eq:interface}) and approximating interfaces as $x_{\pm}^n(t)\approx \pm a_n + \Delta_n(t)$, $n \in \{e,i\}$. 

\begin{figure}[t!]
  \centering
  \includegraphics[width=0.85\textwidth]{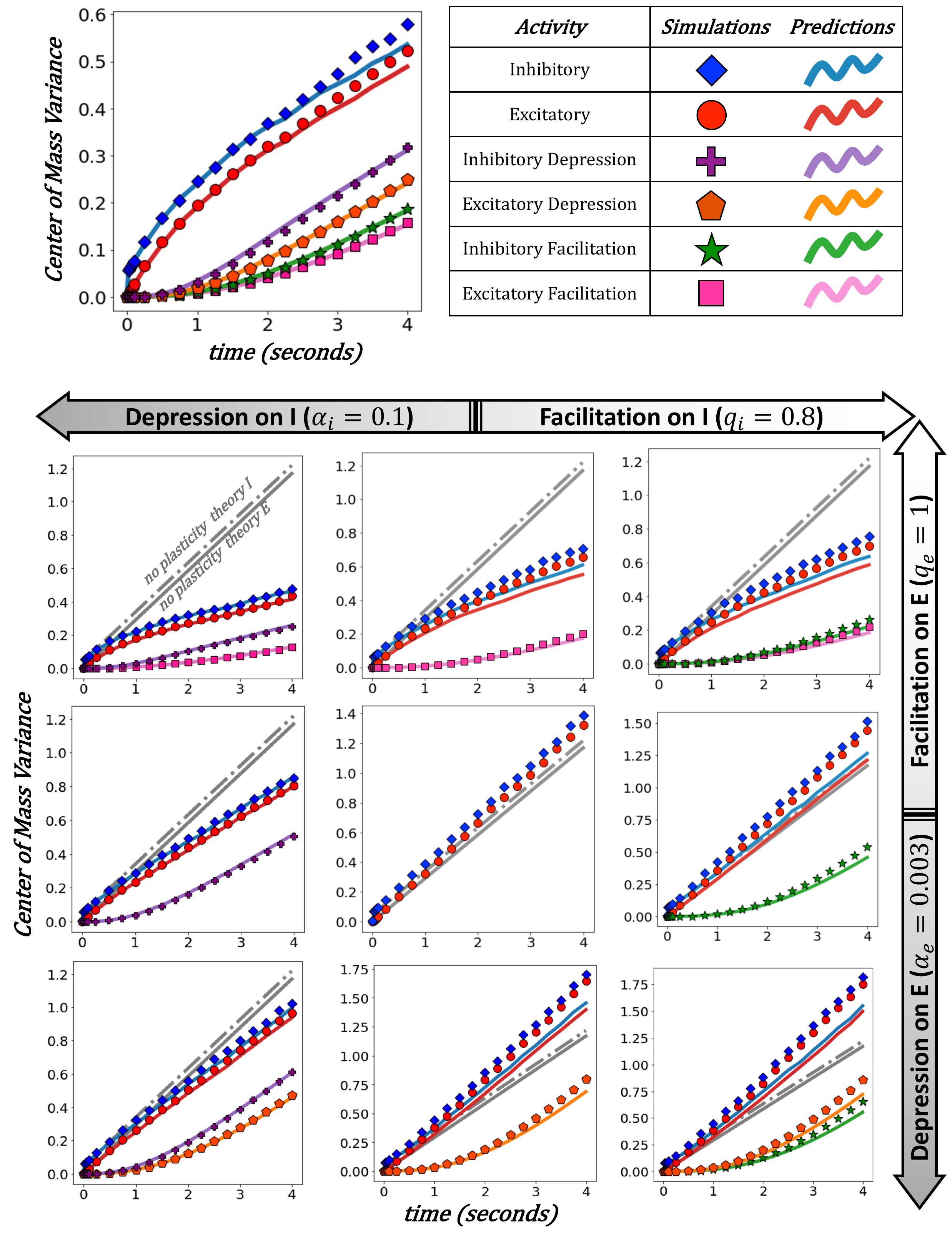}
  \caption{{\bf Center of mass variance can increase sub- or superlinearly depending on the polarity of STP.} (Top): Centroid variances of the neural activity bumps and plasticity profiles as predicted by full simulations (dots) using Eq.~(\ref{eq:model}) and reduced system (lines) using Eq.~(\ref{deln1}) and (\ref{plastcent}). Parameters are as in Table~\ref{tab:params} with $\theta_e=0.4$ and $\theta_i=0.4$, $\tau_{qe}=250$, $\tau_{qi}=200$, $\tau_{re}=150$, $\tau_{ri}=100$, $q_{e0}=1$, $q_{i0}=0.8$, $\alpha_e=0.003$, $\alpha_i=0.1$ (Bottom): Variance trends change as the level of each type of STP is changed according to large grey/white axes (middle black hash mark means zero), growing more slowly (quickly) than an STP-free model given facilitation on E or depression on I (depression on E or facilitation on I). Euler-Maruyama with timestep $dt=0.25$ ms, noise amplitude $\epsilon=0.001$, and truncated spatial interval $x \in [-10,10]$ with steps $dx=0.005$ (for full model) was used to run $10^4$ Monte Carlo simulations for full and reduced systems.}
  \label{fig:8}
\end{figure}

Expanding about the stationary solution and applying the blurring ansatz for $q_{n}$ and $r_{n}$ (and the evolution of $\Delta_{qn,rn}$) developed in Section \ref{s:plastprof} we obtain:
\begin{align}
    d \Delta_n =& \frac{-2(\Delta_e -\Delta_i) {\mc W}_{nm}^-(a_e,a_i) + {\mc P}_{ne}(\Delta_e, \Delta_{re}, \Delta_{qe})-{\mc P}_{ni}(\Delta_i, \Delta_{ri}, \Delta_{qi})}{2 \tau_n {\mathcal G}_n} dt \nonumber \\
    & \hspace{1cm}+ \sqrt{\epsilon \theta_n} d \Xi_n^-(\Delta_n,a_n,t) \label{deln1}
\end{align}
where $m,n \in \{e,i\}$ with $m \neq n$ and
\begin{align*}
 {\mc W}_{nm}^-(x,y) &= w_{nm}(x+y)-w_{nm}(x-y), \\
 {\mc P}_{nm}(\Delta, \Delta_1, \Delta_2) &= \langle r_m(x+\Delta - \Delta_1,t) - 1 + q_m(x + \Delta - \Delta_2,t), {\mc W}_{nm}^-(-a_n,x) \rangle_{a_{um}} \\
 \Xi_n^-(\Delta_n, a_n,t) &= W_n(\Delta_n + a_n,t) - W_n(\Delta - a_n,t),
\end{align*}
where we have defined the inner product over the subdomain $[-a,a]$ as
\begin{align*}
\langle g(x), h(x) \rangle_a = \int_{-a}^a g(x) h(x) dx,
\label{truncip}
\end{align*}
and
\begin{subequations} \label{plastcent}
\begin{align}
    \tau_{qn} d\Delta_{qn}&=-(1+\beta_{qn})(\Delta_{qn}-\Delta_n), \\
\tau_{rn} d\Delta_{rn}&=-\left(1+\frac{1+\beta_n(1+q_{n0})}{1+\beta_n}\alpha_n\right)(\Delta_{rn}-\Delta_n).
\end{align}
\end{subequations}
We validate this reduced nonlinear system describing the coupling between neural and plasticity bump centroids via numerical simulations including uncorrelated noise (\Cref{fig:8}). The slow dynamics of coupling modulated by STP leads to initially nonlinear scalings in variance with time, asymptoting to linear in long time. Facilitation on E (Depression on I) reducing overall variance and Depression on E (Facilitation on I) increasing variance.

Linear approximations can be used to reduce this nonlinear Langevin equation to a multivariate Ornstein-Uhlenbeck process, allowing us to compute covariance and long term diffusion scaling as in \cite{Kilpatrick13FrontiersMultiArea}. Taking the long time ($t \to \infty$) limit of the standard deviation Eq.~\cref{eq:qrplastprof} of the profile deformation, we also have continuity of the blurred plasticity variables for $t>0$, Eq.~(\ref{eq:qrplastprof}), so we can linearize in space:
\begin{align*}
        q_n(x+\Delta) &\approx q_n(x)+(\Delta)\partial_x(q_n(x)) + \dots \\
        r_n(x+\Delta)-1 & \approx r_n(x)-1+(\Delta)\partial_x(r_n(x))+ \dots
\end{align*}
To leading order in the weak noise perturbation, we then have the following pair of linear SDEs with the bump centroids
\begin{align}
    d \Delta_n =& \frac{1}{2 \tau_n {\mathcal G}_n} \left( 2(\Delta_i -\Delta_e) \cdot {\mc W}_{nm}(a_e,a_i) + (\Delta_{qe}-\Delta_e) {\mc Q}_e^n(a_n) \right. \nonumber \\
    & \left. + (\Delta_{re} - \Delta_e) {\mc R}_e^n(a_n) -(\Delta_{qi} - \Delta_i) {\mc Q}_i^n(a_n)-(\Delta_{ri}-\Delta_i) {\mc R}_i^n(a_n) \right) dt \nonumber \\
    & + \sqrt{\ep \theta_n} \left[ dW_n(\Delta_n + a_nt) - dW_n(\Delta_n - a_n,t)\right] \label{delmvou}
\end{align}
for $n,m \in \{e,i\}$ with $m \neq n$, where
\begin{align*}
 {\mc W}_{nm}(x,y) = w_{nm}(x+y)-w_{nm}(x-y), \\
 {\mc Q}_m^n(x) = \int_{-a_{um}}^{a_{um}}(\partial_yq_m(y,t)){\mc W}_{nm}(x,y) dy, \\
 {\mc R}_m^n(x) = \int_{-a_{um}}^{a_{um}}(\partial_yr_m(y,t)){\mc W}_{nm}(x,y) dy,
\end{align*}
along with these linear SDEs for the plasticity variable centroids
\begin{align}
    \tau_{qn} d\Delta_{qn}(t)&=-(1+\beta_{qn})(\Delta_{qn}(t)-\Delta_n(t)), \label{delqmvou} \\
        \tau_{rn} d\Delta_{rn}(t)=&-\left(1+\frac{1+\beta_n(1+q_{n0})}{1+\beta_n}\alpha_n\right)(\Delta_{rn}(t)-\Delta_n(t)). \label{delrmvou}
\end{align}
for $n \in \{e,i\}$. Together, this set of SDEs comprises a closed multivariate Ornstein-Uhlenbeck process:
\begin{equation}
    d \mathbf{\Delta}(t)=\mathbf{M}\mathbf{\Delta}dt+\mathbf{dW}
\end{equation}
where $\mathbf{M}$ is a $6\times 6$ constant matrix consisting of the the coefficients from Eq.~(\ref{delmvou}), (\ref{delqmvou}) \& (\ref{delrmvou}). The covariance matrix can be calculated from the formula
\begin{equation}
    \langle \mathbf{\Delta}(t)\mathbf{\Delta}^T(t)\rangle = \int_0^t \exp{(\mathbf{M}(t-s))}\mathbf{D_M}\exp{(\mathbf{M}^T(t-s))}ds \label{mvoucov}
\end{equation}
where the diffusion matrix has a block structure $\mathbf{D_M} = \left( \begin{array}{cc} \mathbf{D} & \mathbf{0} \\ \mathbf{0} & \mathbf{0} \end{array} \right)$ such that the only nonzero entries are those of the upper left block $\boldsymbol{D}=\begin{pmatrix} D_e&D_c \\ D_c& D_i \end{pmatrix}$
with
\begin{align*}
    &D_n=\frac{\epsilon \theta_n}{2\tau_n^2{\mathcal G}_n^2}[C_n(0)-C_n(2a_n)], \ \ \ n \in \{e, i\}, \\
    \begin{split}
        D_c=\frac{\epsilon \sqrt{\theta_e\theta_i}}{4\tau_e\tau_i{\mathcal G}_e{\mathcal G}_i}[C_c(\Delta_e-\Delta_i+a_e-a_i)-C_c(\Delta_e-\Delta_i+a_e+a_i)\\-C_c(\Delta_e-\Delta_i-a_e-a_i)+C_c(\Delta_e-\Delta_i-a_e+a_i)].
    \end{split}
\end{align*}
\begin{figure}[t!]
  \centering
  \includegraphics[width=0.85\textwidth]{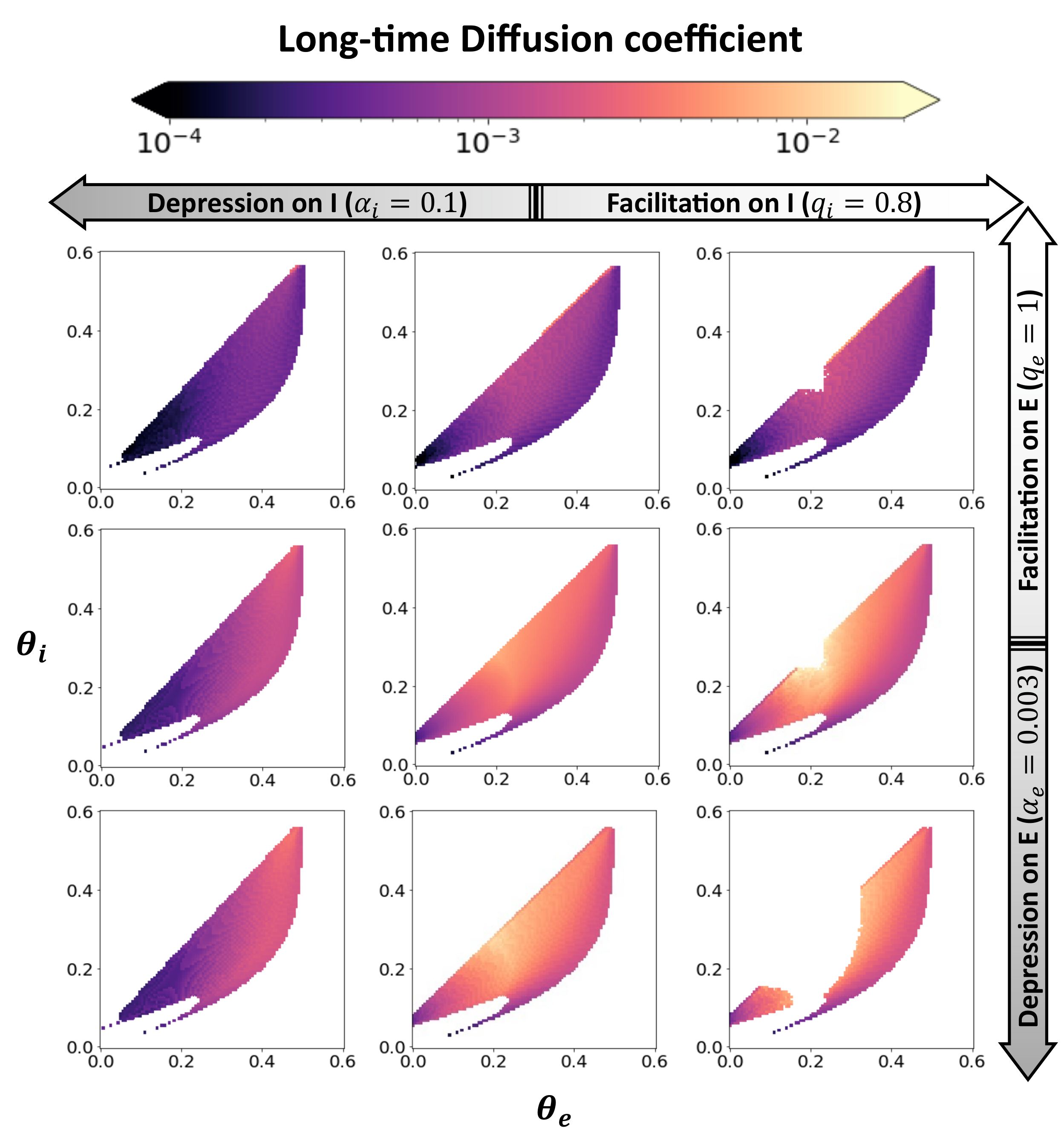}
  \caption{Heatmaps of long-time diffusion coefficients, numerically calculated utilizing Eq.~(\ref{deff}) evaluated at $\mathbf{D}^{\rm eff}_1$ for the long term variance estimated on stable excitatory bump profiles. White areas denote locations of no solutions or unstable solutions. STP parameters are varied according to the large white/grey gradient axes and firing rate thresholds are modified along subaxes $(\theta_e, \theta_i)$. Other parameters are as described in Table~\ref{tab:params} with $t_{end}=20$ seconds}
  \label{fig:9}
\end{figure}
The long time {\em effective} diffusion coefficient can be estimated by averaging the main diagonal elements of the covariance after a fixed time $t_{\rm end}$:
\begin{align}
    \mathbf{D}^{\rm eff} = \frac{{\rm diag}(\langle \mathbf{\Delta}(t_{\rm end}) \mathbf{\Delta}^T(t_{\rm end}) \rangle}{t_{\rm end}} = \lim_{t \to \infty} \frac{{\rm diag}(\langle \mathbf{\Delta}(t) \mathbf{\Delta}^T(t) \rangle}{t}. \label{deff}
\end{align}
The first entry, $\mathbf{D}^{\rm eff}_1$, describes the effective diffusion coefficient of the excitatory neural bump's centroid in long time. Note however, that this effective diffusion approximation typically underestimates the variance of bumps subject to STP since they do not incorporate the initial saturating effects of coupling between centroids. Heatmaps of Eq.~(\ref{deff}) plotted in Figure~\ref{fig:9} exhibit similar trends to our previous results, but exclude the typical sublinear scaling of variance emerging from coupling: Depression on E (Facilitation on I) increases the diffusion coefficient and widen unstable regions, whereas Depression on I (Facilitation on E) lower the diffusion coefficient and shrink unstable regions. Interestingly Depression on I seems to stabilize bumps at lower firing thresholds and has less effect on higher thresholds.

\section{Discussion}
\label{sec:conclusions}
Working memory is a crucial aspect of cognition in humans and other animals~\cite{ma2014changing}. Indeed even simple decisions made from prolonged evidence accumulation may require working memory to retain a running belief formed from information accumulated so far~\cite{waskom2018decision,schapiro2022strategy}. Continuous attractor models in the form of neural field equations provide an ideal balance of biological realism and mathematical tractability for identifying the mechanisms underlying robustness and limitations of working memory~\cite{itskov2011short,Kilpatrick13WandBumpSIAD}. Such approaches therefore define a mathematically and biologically principled link between behavioral observations and the spatiotemporal neural patterns of activity that determine them.

Asymptotic methods for estimating the motion and deformation of wave and pattern solutions to neural field equations are strongly problem dependent due to their nonlinear nature~\cite{bressloff2011spatiotemporal}. One useful approach has been to consider piecewise constant nonlinearities for the firing rate transfer functions~\cite{coombes2004evans}. However, if jumps in these piecewise smooth functions are exposed, standard linear approximations do not hold~\cite{bressloff2011two}. Mollifying jumps using convolutions can rescue this analysis~\cite{kilpatrick2010stability}, but then requires that the sign of perturbations be considered when determining the stability of solutions. We thus aimed to not only resolve how to properly treat jump discontinuities in determining stability of bumps in neural fields with STP, but also in the stochastic dynamics of bumps evolving in response to dynamic fluctuations. This added the interesting feature of nontrivial deformation of STP variables away from their stationary profiles in response to noise and the motion of neural activity variables, requiring an asymptotic theory that could account for {\em blurring}.

Our approach has thus extended standard weak noise approximations to consider the multiple timescale interactions of excitatory/inhibitory neural subpopulations along with synaptic plasticity variables evolving in space-time. The result is a set of nonlinear Langevin equations that not only accounts for the evolution of neural activity and plasticity profiles but also the blurring of plasticity profiles due to the more rapid dynamics of neural activity. This allows us to make accurate approximations of the effect of STP on the variance in bumps over time, which has strong relevance for behavioral predictions in delayed estimation tasks.

An interesting problem that arose here, sidestepped in prior analyses~\cite{kilpatrick2010stability,bressloff2011two}, was the emergence of eigensolutions with complex eigenvalues in the context of our piecewise linear stability analysis. In this case, oscillations will cause the perturbation to change sign at the bump boundaries over time, breaking the fixed sign assumption used to derive the piecewise linear operator. Technically then a simple spectral analysis may  not be sufficient and the transient dynamics may need to be tracked across the sign threshold crossing points. A more thorough analysis along these lines could be to employ switching manifolds that divide the space of perturbation amplitudes for all of the state variables, akin to that recently developed for piecewise linear neural mass models~\cite{coombes2018networks}. Such an approach would much more thoroughly characterize the interaction between the different partitions of the local linearized space as bumps evolve in response to small perturbations.

Another form of perturbation we could consider is weak deterministic stimulation reminiscent of distractors often employed in delayed estimation tasks~\cite{wang2004division}. Our results here suggest that Facilitation on E or Depression on I could help to stabilize bumps in response to such stimulation, so they will not be moved away from their initial position. The detailed perturbation analysis developed here would likely translate, allowing us to insert a spatiotemporal input like $I(x,t)$ in place of the fluctuation terms. Ultimately, using the various forms of asymptotic approximations developed here, we could make behavioral predictions about the level of STP at work in human subjects as a function of their response errors due to distractors~\cite{Seeholzer19}; findings which could be useful in helping to non-invasively identify neuroatypical subjects~\cite{stein2021towards}.



\section*{Acknowledgments}
HLC would like to thank both (a) Sage Shaw for assistance in code development and (b) the noble platypus, whose image inspired her formulation of the blurring kernel approximation and for being the most magnificent monotreme; they never cease to amaze and inspire her.

\bibliographystyle{siamplain}
\bibliography{references}

\appendix
\renewcommand\thefigure{A.\arabic{figure}}   
\renewcommand\thetable{\thesection.\arabic{table}}  

\setcounter{figure}{0} 
\section{Numerical Methods}
\label{app:num}
Numerical simulations and calculations were performed in python. Convolution integrals and the spatially filtered noise were computed via the fast fourier transform convolution method. Time-stepping in full model simulations of Eq.~(\ref{eq:model}) were implemented via the Euler-Maruyama method with initial conditions set to the stationary bumps centered at $x=0$.
Numerical experiments using Milstein method revealed little difference between the two approaches for the parameters tested.
The integrals for the lower dimensional interface predictions Eq.~(\ref{delmvou}) and long time diffusion coefficient Eq.~(\ref{deff}) were approximated as Riemann sums with the spatial step described in corresponding figures. Time and spatial steps are given in corresponding figure captions.

\begin{figure}[t!]
\vspace{-2mm}
  \centering
  \includegraphics[width=0.85\textwidth]{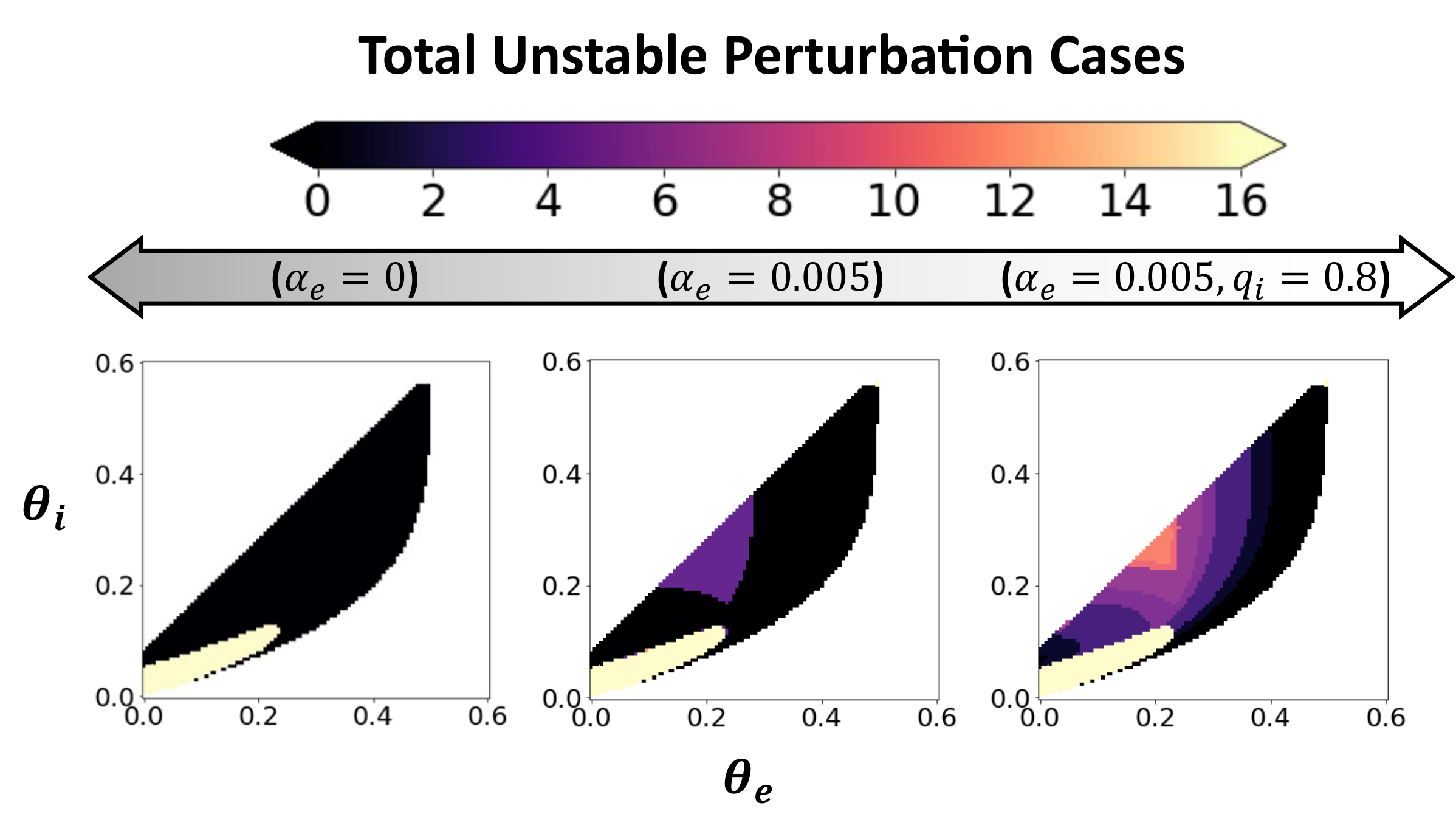}
  \vspace{-7mm}
  \caption{\textbf{Number of sign cases for which there are unstable perturbations and relation to classifying instabilities.} There are sixteen possible cases of perturbation signs (four total edge crossing and two signs each: $2^4=16$). We predict bumps undergo an unstable oscillatory instability if there is a complex eigenpair for each case (off-white region). Drift instabilities account for remaining cases of unstable eigenvalues, requiring at least a single positive real eigenvalue in one of the sign regions (purple regions in center and right panel). The large grey/white gradient axis denotes strength of short term depression $\alpha_e$ and/or facilitation $q_i$. For no plasticity, oscillatory instabilities emerge as in \cite{cihak2022distinct}. Overlapping regions with drift and oscillatory instabilities are classified by examining the class of the dominant eigenvalue.}  
  \vspace{-2mm}
  \label{fig:unstableclass}
\end{figure}

\begin{figure}[b!]
\vspace{-2mm}
  \centering
  \includegraphics[width=0.85\textwidth]{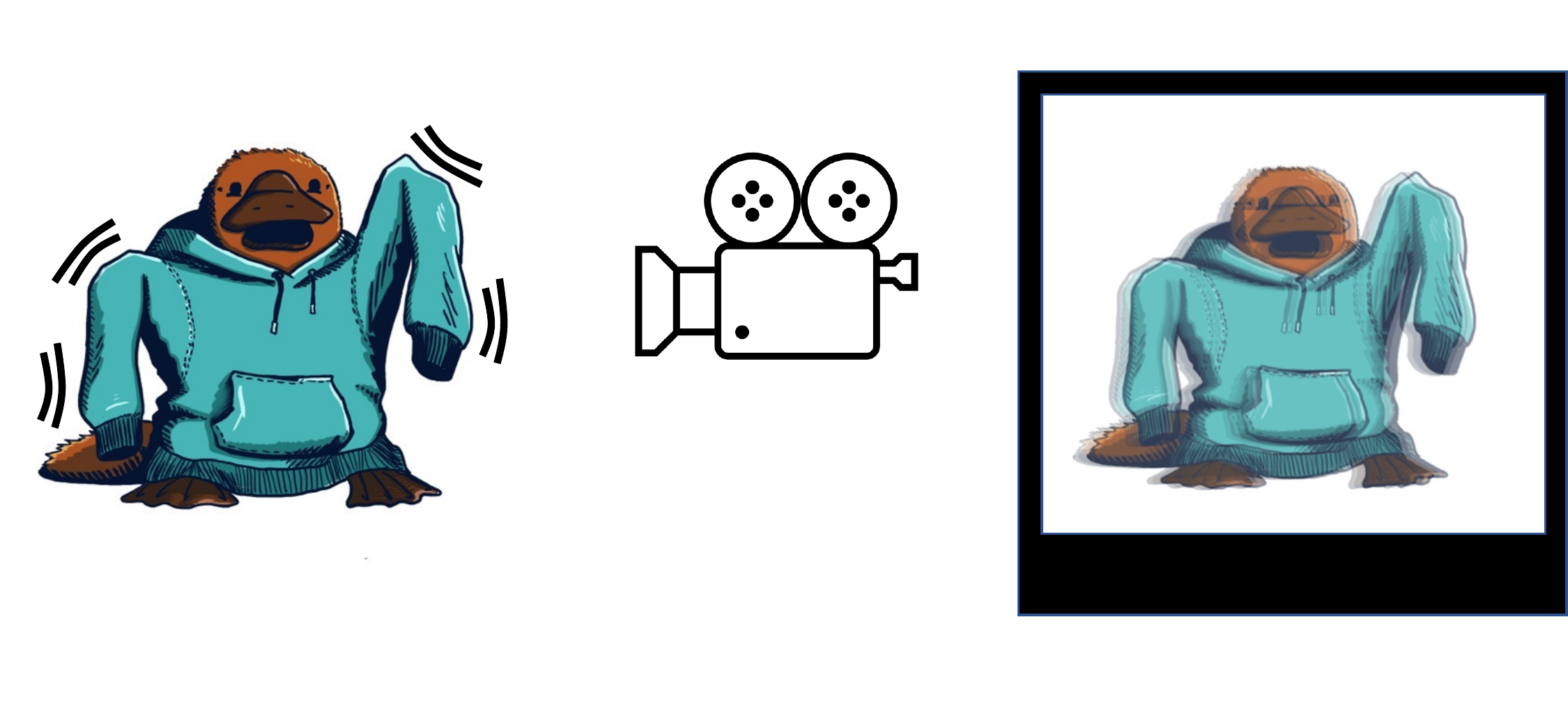}
  \vspace{-7mm}
  \caption{\textbf{Motion blur} Pictures of moving objects taken with cameras at slow shutter speed would result in blurred images. Illustration made by HLC.}  
  \vspace{-2mm}
  \label{fig:platypusBlur}
\end{figure}

\section{Classifying unstable solutions} \label{app:unst} Our stability analysis relies on an approach originally developed in \cite{kilpatrick2010stability,bressloff2011two} which first mollifies jumps and exposed singularities emerging from linearizing steps in plasticity profiles. Thereafter, a standard spectral theory of stability is not possible, but must be modified to account for the sign of perturbations at bump boundaries. Eigensolutions are then found by considering all possible sign combinations and solving for eigenfunctions and eigenvalues. If any real eigenvalues are positive, then the underlying solution being linearized around is unstable, implying either a drift or collapse/expansion instability. If not, but there are complex eigenvalues with positive real part, we conjecture solutions will also be unstable. The eagle-eyed reader may note that perturbation signs will change given an oscillatory instability, but if each perturbation sign case has associated positive real part complex eigenvalues pairs, the perturbation will not decay and indeed the bump will be unstable (e.g., long region protruding from the lower left corner of phase space in \Cref{fig:unstableclass}). There can also be cases where both drift and oscillatory instabilities can be present. 

\section{Motion Blur} \label{app:blur}
Shifts in the neural E/I active region cause relaxations in the plasticity variables so they evolve as an {\em image} of the active regions, generating smearing analogous to visual phenomena emergent from cameras with slow shutter speed (\Cref{fig:platypusBlur}). Plasticity profiles therefore reflect probabilities of where active regions may be. Such {\rm image motion blur} provides a rough estimate of the most likely locations of the active region. Motivated by models of motion blur from image processing~\cite{smith1998characterization, ji2008motion, solomon2011fundamentals}, we inferred an ansatz for the blurring kernel, applied to generate the observed diffusion of the initial plasticity profiles. Our heuristically estimated blur kernel provides accurate descriptions across a broad parametric range of simulation.

\end{document}